\documentclass[a4paper,11pt]{article}
\usepackage{soul}
\usepackage[usenames,dvipsnames]{color}
\usepackage{jheppub}
\usepackage{esvect}
\usepackage{amsmath, amssymb, slashed, epsf, color, graphicx, latexsym, tensor, physics, xcolor}
\usepackage{epsfig}
\usepackage{comment}
\usepackage{graphics}
\usepackage{mathtools}
\usepackage{appendix}

\title{Revisiting Leading Quantum Corrections to Near Extremal Black Hole Thermodynamics}
\author{Nabamita Banerjee and}
\author{Muktajyoti Saha}

\affiliation{Indian Institute of Science Education and Research Bhopal,\\ Bhopal Bypass, Bhauri, Bhopal 462066, India}

\emailAdd{nabamita@iiserb.ac.in} 
\emailAdd{muktajyoti17@iiserb.ac.in}

\abstract{Computing the 4D Euclidean path integral to one-loop order we find the large quantum corrections that govern the behavior of a spherically symmetric non-supersymmetric near-extremal black hole at very low temperature. These corrections appear from the near-horizon geometry of the near-extremal black hole. Using first-order perturbation theory we find that such corrections arise from the zero modes of the extremal background. In the logarithm of the partition function, these correspond to terms involving logarithm of temperature. Part of our result matches with the existing one in literature derived from an effective  Schwarzian theory. }

\begin{document}

\maketitle
\section{Introduction}

Black holes are thermal objects, uniquely described in the General Theory of Relativity by their mass, angular momentum, and charges. A revolutionary discovery in physics is the understanding of the laws of black hole thermodynamics, where the temperature is given by the surface gravity and the entropy is given by the area of the horizon \cite{Bekenstein:1973ur, Hawking:1976de} of the black hole. In \cite{Gibbons:1976ue, York:1986it}, it has been shown that the entropy of a black hole can be computed from a semiclassical computation of the Euclidean path integral in the black hole background. Later in \cite{Wald:1993nt}, it was shown that the area law of entropy for a black hole with non-vanishing temperature can also be obtained as the Noether charge corresponding to the time translation Killing vector, evaluated on the black hole horizon. Beyond the semiclassical regime, the entropy gets universal corrections of the form of logarithm of horizon area \cite{Solodukhin:1994yz, Mann:1997hm, Medved:2004eh, Cai:2009ua, Aros:2010jb}. Like ordinary thermodynamic systems, black hole entropy should also have a microscopic description in terms of the degeneracy of states in quantum theory. For a certain class of charged black holes, namely extremal black holes, the microscopic counting is very well understood in the context of string theory \cite{Strominger:1996sh, Breckenridge:1996is, David:2006ru, David:2006ud, Sen:2008yk, Gupta:2008ki, Banerjee:2007ub, Banerjee:2008ky, Banerjee:2009uk, Banerjee:2009af}.

A charged black hole at nonzero temperature, called a non-extremal black hole, has two distinct horizons.  
Such a non-extremal black hole emits thermal radiation \cite{Hawking:1975vcx, Wald:1975kc} and eventually settles to the ground state which corresponds to the extremal black hole.
An extremal black hole is a charged black hole at zero temperature for which the two horizons coincide. For these black holes,  Wald's formalism for computing entropy does not apply. Sen in \cite{Sen:2005wa, Sen:2007qy} computed their entropy using the entropy function formalism and obtained the correct area law, see also \cite{Sahoo:2006pm, Sahoo:2006rp}. It was shown that an extremal black hole has an infinitely long AdS$_2$ throat near the horizon which results in an enhanced isometry. This is particularly important in understanding the dynamics of these black holes. Going beyond the semiclassical limit, in \cite{Banerjee:2010qc, Banerjee:2011jp, Sen:2012cj} the logarithmic corrections were computed for extremal black holes and agreement with microscopic results in several scenarios was established. Clearly, extremal black holes play a very important role in understanding the microstructure of black holes. The logarithmic terms in black hole entropy were also computed in various other cases \cite{Sen:2012kpz, Ferrara:2011qf, Sen:2012dw, Bhattacharyya:2012wz, Karan:2017txu, Karan:2019gyn, Karan:2020njm, Karan:2021teq, Banerjee:2021pdy, Karan:2022dfy}, although the microscopic results are not available for such systems. These logarithmic corrections do not depend on the explicit ultraviolet structure of the underlying quantum theory of gravity. Rather, these are generated by loops of massless fields present in the theory. These corrections are universal features of the theory that can be extracted from the infrared data and yet these are very important to constrain the UV-complete theories.

For non-extremal black holes, a concrete microscopic understanding is so far lacking. This puts the study of near-extremal black holes on a very important footing. 
They can be considered as small temperature deviations from the extremal black holes ones to enjoy the reminiscence of that arise at extremality and simultaneously correspond to excited states on the microscopic side. 
 On the macroscopic side, a naive semiclassical analysis for a near-extremal black hole gives the energy above extremality to be proportional to the square of temperature. However, the average energy of Hawking quanta is proportional to temperature. This seems to suggest that at sufficiently low temperature, the near-extremal black hole does not have enough energy to radiate, which is a clear contradiction to the concept of Hawking radiation. As a resolution to the apparent puzzle, in \cite{Iliesiu:2020qvm}, it was argued that semiclassical physics breaks down at such small temperatures and to understand the system, one needs to incorporate quantum corrections to the thermodynamics. The authors considered the effective description \cite{Nayak:2018qej, Moitra:2019bub, Iliesiu:2020qvm} of the near-extremal black holes, where the low energy physics is described by a Schwarzian theory of slightly broken asymptotic symmetry modes of the AdS$_2$ factor of extremal near-horizon geometry. Using the path integral of Schwarzian theory \cite{Stanford:2017thb, Saad:2019lba}, a large quantum correction of the form $\log{T}$ appears in the logarithm of the partition function. These corrections are different than the logarithm of horizon area (or charge) correction although both of these come from the one-loop computation. Using the $\log{T}$ term, the average energy expression gets an extra contribution that resolves the apparent contradiction involving Hawking radiation. This is because, in presence of this correction the average black hole energy remains greater than that of the Hawking quanta even at very low temperatures.

In this paper, we attempt to extract the $\log{T}$ correction from a direct 4D Euclidean path integral computation without resorting to the effective lower-dimensional description. We observe that these corrections cannot be obtained by taking a small temperature limit of the results for a non-extremal black hole. 
Instead, we carry on the analysis in a limit where the near-extremal solution is treated as a small deviation of the extremal solution. The computation of the partition function for an extremal background is completely captured by the infinite near-horizon throat. Although the throat is finite for a near-extremal black hole, it is very large as the  temperature is small. In the asymptotic region, the geometry is well-approximated by the full extremal solution. Here the effects of temperature are highly suppressed. Since the fluctuations die off near asymptotic infinity, the quantum corrections near the horizon have a more significant contribution than that in the asymptotic region. Hence, even in this case, the dynamics is governed by the near-horizon data. In this spirit, we quantize the system in the near-horizon region of the near-extremal black holes.

The computation of one-loop partition function amounts to evaluating the eigenvalues of the kinetic operator corresponding to small fluctuations around a background. Since the near-horizon near-extremal background is a deviation from the extremal AdS$_2\times$S$^2$ geometry, the near-extremal kinetic operator is a small temperature deviation of the extremal kinetic operator. The eigenfunctions of the extremal kinetic operator are known which allows us to employ the first-order perturbation theory technique to find the near-extremal eigenvalues. We notice that the $\log{T}$ correction generates from the zero modes of the extremal kinetic operator, which get a small non-zero mass due to the near-extremal correction of the background. All other modes give rise to contributions, polynomially suppressed in temperature. Therefore, we find the zero modes of the extremal kinetic operator and compute the corresponding eigenvalue corrections. The $\log{T}$ correction coming from the tensor zero modes (asymptotic symmetries of AdS$_2$), is in agreement with the Schwarzian results. However, we get additional corrections from other zero modes. Finally, we would like to comment that the issues raised in this paper are similar in spirit to that of \cite{Iliesiu:2022onk}, but the explicit analysis and computations are different. Also, we differ in our interpretation of the results. 

The paper is organized as follows: In section \ref{NE-semicl-sec} we discuss the near-horizon geometry of a near-extremal black hole in 4D Einstein-Maxwell theory and compute the Bekenstein-Hawking entropy from the near-horizon geometry only. This signifies that at least at the semiclassical level, the near-horizon information is enough to find the entropy of the system. In section \ref{NE-quant-sec}, we discuss the forms of the quantum correction to near-extremal partition function and lay out our strategy of computing $\log{T}$ contributions. Using first-order perturbation theory, we compute the $\log{T}$ corrections in section \ref{NE-logT-sec}. In section \ref{1D-sec}, we present an effective Schwarzian description that captures part of the 4D computations. Finally, we summarize the results in section \ref{summary-sec}. The appendices contain some relevant computational details.

\section{Near-extremal black hole in 4D Einstein-Maxwell theory} \label{NE-semicl-sec}
We consider the 4D Einstein-Maxwell action in Euclidean signature:
\begin{align}
    \mathcal{S} = -\frac{1}{16\pi G_N}\int d^4x \sqrt{g} (R - F^2) \label{EM-act}.
\end{align}
We will set $16\pi G_N = 1$ for convenience. The Euclidean time direction is compact. For a well-defined variational problem, we add appropriate boundary terms near asymptotic infinity in the spatial direction. Imposing Dirichlet and Neumann boundary conditions on the metric and gauge field respectively, the required boundary term \cite{Gibbons:1976ue, York:1986it, PhysRevD.42.3376, Hawking:1995ap} is given by,
\begin{align}
    \mathcal{S}_{\text{bdy}} = -2\int \sqrt{\gamma}(K + 2n_A A_B F^{AB}), \label{EM-bdy}
\end{align}
here $\gamma$ is the induced metric and $n_A$ is the outward normal to the boundary. Varying the action \eqref{EM-act} along with the boundary terms, we have the equations of motion given as:
\begin{align}
   & R_{AB} = 2 F_{AC}\tensor{F}{_B^C} - \frac{1}{2}g_{AB}\bar{F}^2; \qquad R = 0; \qquad \nabla_A F^{AB} = 0. \label{eom}
\end{align}
The classical solutions satisfy these equations of motion and also the Bianchi identities, given by,
\begin{align}
    & \nabla_{[A} F_{BC]} = 0; \qquad R_{A[BCD]} = 0. \label{bianchi}
\end{align}
Spherically symmetric black hole solutions in this theory are given by Reissner-N\"{o}rdstrom geometry, labeled by mass and charge parameters. For a black hole solution, the periodicity of the time direction is fixed by the inverse temperature. We are interested in a near-extremal black hole solution that has a very small temperature. This solution is perturbatively close to the zero-temperature extremal solution. We will now briefly discuss the geometries.

\subsection {The full extremal solution and its near horizon geometry}
In this subsection, we will discuss the extremal Reissner-N\"{o}rdstrom solution since we will be treating the near-extremal solution as a small deviation from extremality. We begin with the generic non-extremal Reissner-N\"{o}rdstrom solution\footnote{Without loss of generality we are considering electric charge only since in 4D, we have electric-magnetic duality.} in the theory \eqref{EM-act}, 
\begin{align}
   & ds^2 = g_{AB}dx^A dx^B = f(r)dt^2 + \frac{dr^2}{f(r)} + r^2d\Omega^2, \quad f(r) = 1 - \frac{2M}{r} + \frac{Q^2}{r^2}, \\
   & A_t = i Q \left( \frac{1}{r_{+}} - \frac{1}{r} \right) , \quad F_{rt} = \frac{i Q}{r^2}.  \label{RN}
\end{align}
This solution has two horizons\footnote{We note that the two horizons are visible in the Lorentzian geometry. The Euclidean geometry starts from $r = r_{+}$, while the time direction has periodicity equal to the inverse temperature.} at $r_{\pm} = M \pm \sqrt{M^2 - Q^2}$ such that $f(r_{\pm}) = 0$. It is preferable to write the solution in terms of the parameters $Q$ and $r_{+}$ for the discussion of near-extremal black holes. We have the following relations,
\begin{align}
    M = \frac{1}{2r_{+}}(Q^2 + r_{+}^2), \quad r_{-} = \frac{Q^2}{r_{+}}.
\end{align}
The temperature is given by,
\begin{align}
    T = \frac{1}{4\pi}\abs{f'(r_{+})} = \frac{1}{4\pi r_{+}^3}(r_{+}^2 - Q^2).
\end{align}
At extremality, the two horizons coincide such that $M=Q=r_0$, where $r=r_0$ denotes the extremal horizon. For the extremal black hole, $f(r_0) = 0$ and $f'(r_0) = 0$. Then the $g_{tt}$ component of the metric takes the following form which now has a double zero at $r=r_0$,
\begin{align}
    g_{tt} = f(r) = \left(1-\frac{r_0}{r}\right)^2.
\end{align}
In the near-horizon region i.e. for $r-r_0 = \rho \ll r_0$, the solution can be expressed as,
\begin{align}
    ds^2 = \frac{\rho^2}{r_0^2}dt^2 + \frac{r_0^2 d\rho^2}{\rho^2} + r_0^2 d\Omega^2, \quad F_{rt} = \frac{i}{r_0}.
\end{align}
Therefore the geometry is AdS$_2\times$S$^2$ near the horizon. In this region, the symmetry gets enhanced due to the AdS$_2$ factor which plays a very important role in the dynamics of these black holes.

\subsection{The full near-extremal solution and its near horizon geometry} \label{NHR-sec}
Next, keeping the charge fixed to its extremal value $r_0$, we introduce a small mass above extremality such that the black hole becomes slightly non-extremal. 
 As a consequence, the near-horizon geometry of a near-extremal black hole is described by a small deviation from AdS$_2\times$S$^2$. Before moving ahead with the explicit structure of the geometry, let us briefly mention the effective 2D description of the near-horizon physics of such black holes, as presented in the existing literature \cite{Nayak:2018qej, Moitra:2019bub, Iliesiu:2020qvm}. Using the symmetries of the near-horizon region, the 4D theory can be reduced to a two-dimensional manifold which, in the massless sector, gives rise to a 2D theory of gravity coupled to dilaton. An appropriate Weyl transformation of the 2D metric removes the kinetic term of the dilaton. The constant dilaton solution in this theory corresponds to the near-horizon extremal geometry. The standard procedure to describe near-extremal physics is to consider fluctuations of only the dilaton field around its constant value, while keeping the metric part same. At first order in fluctuations, the resulting theory turns out to be Jackiw-Teitelboim (JT) gravity \footnote{JT is a 2D gravitational theory, coupled to a dilaton, described by the action: $$-\frac{1}{16\pi G_2}\int d^2x \sqrt{g}\phi (R+2) - \frac{1}{8\pi G_2}\int dx\sqrt{\gamma}\phi K.$$}, with appropriate boundary conditions \cite{Jackiw:1984je, Teitelboim:1983ux}. By integrating out the dilaton, JT gravity can be further boiled down to a 1D Schwarzian theory \cite{Maldacena:2016upp, Saad:2019lba}, which captures the near-extremal physics. This puts a constraint on the 2D metric, which sets the curvature to a negative constant value i.e. the metric is fixed to asymptotically AdS$_2$. The falloff of the dilaton also gets fixed near the boundary. Thus the effective JT description suggests that the near-horizon geometry of the near-extremal black hole is a Weyl transformed  AdS$_2$, where the conformal factor is fixed by the dilaton profile with a sphere, having a slightly varying radius, also given by the dilaton. This form of the solution is however critical, since it does not solve the 4D equations of motion. In this section, we directly compute the near-horizon geometry from 4D Reissner-N\"{o}rdstrom solution, which also satisfies the equations of motion to leading order in deviation from extremality. We argue that this near-horizon geometry (after considering suitable Weyl factor) cannot be transformed into a locally AdS$_2$ geometry and hence is not equivalent to the solution coming from JT gravity. Our effective description of the system is presented in section \ref{1D-sec}.

We now present the near-extremal geometry. Due to the presence of a small temperature, the horizons split slightly from the extremal one. We  parametrize the near-extremal solution by $r_0$ and $\delta$, where $\delta \ll r_0$ characterizes the first-order deviation from extremality\footnote{Since $\delta\sim T$, we will use the temperature $T$ as the perturbation parameter in the computation of one-loop determinant so that we can directly extract out the $\log{T}$ dependence. But for the semiclassical computation from the near-horizon geometry, it is instructive to parametrize the solution by $\delta$.}. In terms of these parameters we have,
\begin{align}
    & M = r_0 + \frac{\delta^2}{2r_0} + \frac{2\delta^3}{r_0^2} + \mathcal{O}(\delta^4), \nonumber\\
    & r_{+} = r_0 + \delta + \frac{5\delta^2}{2r_0} + \mathcal{O}(\delta^3), \nonumber \\
    & T = \frac{\delta}{2\pi r_0^2} + \mathcal{O}(\delta^3), \quad \beta = \frac{2\pi r_0^2}{\delta} + 16\pi\delta - \frac{45\pi\delta^2}{2r_0} + \mathcal{O}(\delta^3). \label{NE-parameters}
\end{align}

Hence, the full near-extremal solution gets corrected at order $\delta^2$. It is given by \eqref{RN} with the $g_{tt}$ component being,
\begin{align}
f(r) = \left(1-\frac{r_0}{r}\right)^2 - \frac{\delta^2}{r r_0}.
\end{align}
We will split the full near-extremal solution into near-region and far-region, which will be important for the computations. From effective 2D perspective, such a splitting was performed in \cite{Nayak:2018qej, Moitra:2019bub, Iliesiu:2020qvm}. 

\subsubsection*{The geometry in near-horizon region (NHR):} 
First, we consider the near-horizon geometry of the near-extremal RN black holes. We perform the following coordinate transformations on the RN geometry \eqref{RN} with parameters \eqref{NE-parameters},
\begin{align}
    r(\eta) = r_{+} + \delta(\cosh{\eta}-1), \quad t(\theta) = \frac{r_0^2}{\delta}\theta, \label{NH-coord}
\end{align}
where, the coordinates range from $0<\eta<\eta_0$ and $0<\theta<2\pi$.  We denote the coordinates on AdS$_2$ by $x^\mu$ and the coordinates on S$^2$ by $x^i$. The horizon is located at $\eta = 0$, such that $r = r_{+}$. In this coordinate system, the near-extremal geometry has the form $ { 
\tilde g_{AB} = {g}^0_{AB} + \delta g^{(c)}_{AB}, 
\tilde F_{AB} = {F}^0_{AB} + \delta F^{(c)}_{AB}, \tilde A_B = {A}^0_B + \delta A^{(c)}_B }$ where\footnote{The same can be obtained by perturbatively solving the 4D equations of motion directly in the near-horizon region as illustrated in appendix \ref{NH-geo-app}.}
\begin{align}
     & {g}^0_{AB}dx^A dx^B = r_0^2 (d\eta^2 + \sinh^2{\eta}d\theta^2) + r_0^2 (d\psi^2 + \sin^2{\psi}d\varphi^2), \nonumber \\
     & {F}^0_{\mu\nu} = \frac{i}{r_0}\varepsilon_{\mu\nu}, \quad {A}^0_{\theta} = i r_0 (\cosh{\eta} - 1). \label{ext-NH} 
\end{align}
These are the $\mathcal{O}(1)$ pieces of the expansion that give the near-horizon extremal geometry. Note that at this order, the horizon is located at $\eta = 0$ or at $r = r_0$, which is the extremal horizon. The $\mathcal{O}(\delta)$ correction is given as,
\begin{align}
     & g^{(c)}_{AB}dx^A dx^B = 2 r_0 (2+\cosh{\eta})\tanh^2\left(\frac{\eta}{2}\right) (d\eta^2 - \sinh^2{\eta}d\theta^2) + 2 r_0 \cosh{\eta} d\Omega^2, \nonumber\\
     & F^{(c)}_{\mu\nu} = - 2 i r_0^{-2}\cosh{\eta}\varepsilon_{\mu\nu}, \quad  A^{(c)}_{\theta} = -i\sinh^2{\eta}. \label{NE-corr}
\end{align}
Here the perturbative parameter is the small deviation of horizon $\delta$, proportional to the temperature. $\varepsilon_{\mu\nu}$ is the Levi-Civita tensor on AdS$_2$, with the non-zero component being ${\varepsilon_{\eta\theta} = r_0^2 \sinh{\eta}}$. This geometry has also been discussed in \cite{Iliesiu:2022onk}. Two important points to note are, 
\begin{itemize}
    \item  We are considering a near-extremal black hole with a very small temperature $T$, so that we have $\delta\ll r_0$ or $r_0 T \ll 1$. The perturbative expansion of the near-horizon geometry is valid as long as we are very close to the horizon so that the new radial coordinate $\eta$ does not grow much. Hence, we choose the radial cutoff $\eta_0$ such that $\delta\text{e}^{\eta_0} \ll r_0$. For an extremal black hole, this radial cutoff can be taken to infinity, resulting in an infinite AdS$_2$ throat.

    \item  From the structure of the near-extremal correction, we note that the geometry on the $(\eta, \theta)$ plane is not asymptotically AdS$_2$. All the corrections to the fields appear at the same order of temperature and they diverge near the cutoff surface at $\eta = \eta_0$. Since the deviation $g^{(c)}_{\mu\nu}$ is traceless with respect to the AdS$_2$ metric, it cannot be transformed to even a small Weyl transformation of AdS$_2$ via coordinate transformations. This point is in contradiction with a 2D effective description of these black holes in terms of a JT-like theory, since, for JT theory, the background must be a locally AdS$_2$ geometry. We shall expand on this in the discussion section. 
\end{itemize}

\begin{figure}[h]
\centering
\includegraphics[width=0.7\textwidth]{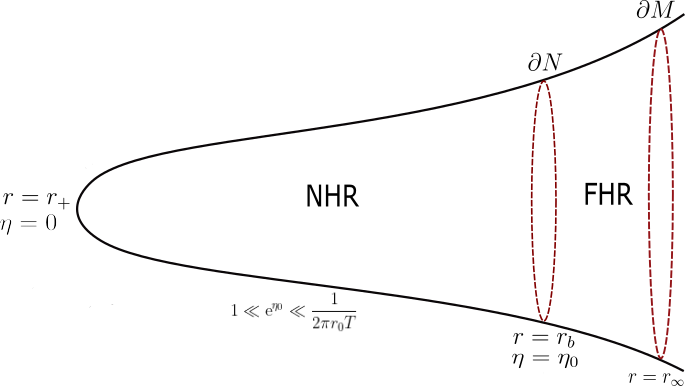}
\caption{Splitting of the geometry into near-horizon and far-horizon regions}
\end{figure}

\subsubsection*{The geometry in far-horizon region (FHR):}
In the far region, we need to consider the full solution, where the corrections appear at $\mathcal{O}(\delta^2)$. At large enough distances from the horizon, the geometry closely resembles the full extremal geometry as the horizons appear to be overlapping. Hence in the FHR, the effects of temperature become negligible as compared to that in the NHR. 

So far we have split the full near-extremal geometry into near-horizon and far-horizon regions. These regions are separated by a 3D boundary curve located at $\eta = \eta_0$ or $r = r_b$. We denote the boundary as $\partial N$. The parameters $\eta_0$ and $r_b$ are related through the coordinate transformation \eqref{NH-coord}. The fields are smooth across this artificial boundary. We impose Dirichlet boundary condition on the metric and Neumann boundary condition on the gauge field. Physically these two conditions fix the length of the boundary and the charge of the black hole respectively.

To summarize, the full manifold ($M$) is obtained by gluing the two geometries across $\partial N$. The NHR manifold has a boundary $\partial N$ whereas the FHR manifold has two boundaries $\partial N$ and $\partial M$. The near-horizon boundary $\partial N$ is shared by both the manifolds and $\partial M$ is the boundary located near asymptotic infinity. We will work in a limit such that the boundary $\partial N$ is asymptotically far from the horizon with respect to the NHR but it still lies in the near-horizon region with respect to asymptotic infinity. These limits also have been discussed in \cite{Nayak:2018qej, Moitra:2019bub, Iliesiu:2020qvm} and are given in equations \eqref{NH-rad-cutoff}.

\subsection {Semiclassical near-extremal entropy from near-horizon geometry} \label{ent-NHR-sec}

The thermodynamics of the near-extremal black hole can be studied using the full geometry as discussed in appendix \ref{RN-thermo-app}, where we work in an ensemble with fixed charge and fixed length of the boundary at asymptotic infinity. In this section, we will extract the Bekenstein-Hawking entropy from the near-horizon region only without referring to the far-horizon data. This is because entropy is a near-horizon quantity for any black hole, which can be anticipated from Wald's derivation of entropy as the Noether charge at horizon \cite{Wald:1993nt}. For the computation of entropy, we don't need additional counterterms \cite{York:1986it}, since the role of counterterms is only to regulate the energy via appropriate background subtraction. For computing the entropy, we need to consider the boundary length as an independent parameter for our choice of ensemble. This plays the role of the inverse temperature from the perspective of an observer in the near-horizon boundary. For this purpose, we need to parametrize the black hole solution with charge $Q = r_0$ and the shift $\delta$ in the horizon radius (or mass above extremality) instead of parametrizing by temperature, which gives the boundary length near asymptotic infinity.

The near-horizon geometry, that describes the small-temperature physics above extremality, 
 has been discussed in section \ref{NHR-sec}. This geometry, given by \eqref{ext-NH} and \eqref{NE-corr}, is well-approximated to describe the same up to a radial distance $\eta_0$ such that $\eta_0$ is large but the near-extremal corrections (terms proportional to $\delta$) remain small compared to the extremal geometry. Therefore we have,
\begin{align}
    & \delta \text{e}^{\eta_{0}}\ll r_0, \\
    & \text{e}^{\eta_{0}} \approx \frac{r_0}{\delta}\epsilon \label{NH-rad-cutoff}, \quad \epsilon \ll 1.
\end{align}
To get the entropy, We evaluate the action \eqref{EM-act} along with the boundary terms \eqref{EM-bdy} for the near-horizon near-extremal solution, where the boundary is located at radial distance $\eta = \eta_0$ in the NHR. The on-shell action is given as,
\begin{align}
    & I = 16\pi(-\pi r_0^2 - 2\pi r_0\delta\cosh^2{\eta_0}).
\end{align}
The boundary length is given as,
\begin{align}
    \beta_0 = \frac{1}{r_0}\int_{\eta = \eta_0} d\theta \sqrt{g_{\theta\theta}} = 2\pi\sinh{\eta_0} - \frac{2\pi\delta}{r_0}\csch{\eta_0}(2 - 3\cosh{\eta_0}+\cosh^3{\eta_0}).
\end{align}
Now we use the condition \eqref{NH-rad-cutoff} so that the near-horizon approximation holds and we work in small $\epsilon$ limit. The entropy is given by,
\begin{align}
     S_{\text{near-ext}} = \beta_0 \frac{\partial I}{\partial \beta_0} - I = \beta_0 \frac{\partial I}{\partial \delta}\frac{\partial \delta}{\partial\beta_0} - I = 16\pi^2 r_0^2\left(1+\frac{2\delta}{r_0}\right).
\end{align}
This result is obtained for small $\delta$ and $\epsilon$ and it is equal to horizon area to linear order in $\delta$. In terms of the temperature parameter, we recover the semiclassical entropy as:
\begin{align}
     S_{\text{near-ext}} = 16\pi^2 r_0^2\left(1+4\pi r_0 T\right).
\end{align}

Therefore, we see that the Wald entropy \cite{Wald:1993nt} can be independently computed from the near-horizon geometry only. The result is of course in agreement with the computation using full geometry as presented in appendix \ref{RN-thermo-app}, where we also discuss the computation of energy. In the subsequent sections, we compute the quantum $\log T$ correction to the semiclassical result, which is the main goal of this paper.

\section{Quantum corrections to near extremal black hole partition function} \label{NE-quant-sec}

The contribution to entropy coming from terms proportional to the logarithm of area has been a subject of huge interest in the context of extremal and non-extremal black holes \cite{Banerjee:2010qc, Banerjee:2011jp, Sen:2012kpz, Sen:2012cj, Sen:2012dw}. This appears from the total one-loop correction to the partition function due to the presence of massless fields. On one hand, these corrections can be computed from the low energy data i.e. the computations do not require the ultraviolet information of the underlying quantum theory. On the other hand, the universal feature of these log corrections allows more control over the microstructure of the black holes. For certain classes of extremal black holes, these corrections match with the microscopic results \cite{Banerjee:2010qc, Banerjee:2011jp, Sen:2012kpz, Sen:2012cj}. A similar study for near-extremal black holes is also very important, as these systems can be considered as small temperature deviations from extremal black holes. Furthermore, at very low temperatures the semiclassical thermodynamic description is not enough to study the dynamics of these black holes \cite{Iliesiu:2020qvm}, as we describe below. 

\subsection{Breakdown of semiclassical physics}

As noted in the introduction, the semiclassical analysis breaks down at sufficiently low temperature. Let us briefly discuss the importance of quantum corrections for a near-extremal black hole at very low temperatures. It can be understood from the expression of mass \eqref{NE-parameters}, which is proportional to the energy of the system \eqref{NE-energy} under semiclassical approximation. In terms of temperature, it is given as,
\begin{align}
    E = 16\pi(r_0 + 2\pi^2r_0^3 T^2).
\end{align}
Therefore, the thermodynamic energy above extremality goes as $\sim T^2$. But this is inconsistent with Hawking radiation since the average energy of thermal radiation goes as $\sim T$. Below a certain mass scale $M_{\text{gap}}\sim r_0^{-3}$, the semiclassical energy of the black hole is less than that of the average energy of radiation. This implies that the black hole cannot radiate even though it has a nonzero temperature.  To resolve this issue it was conjectured that there exists a literal mass gap of order $M_{\text{gap}}$ between the extremal and lightest near-extremal states, although in a non-supersymmetric theory, the rationale of the gap was not justified and hence the conjecture is critical. A resolution was proposed in \cite{Iliesiu:2020qvm}, where the authors argued that, at very low temperatures, semiclassical description breaks down and one has to take quantum effects into account. They further used a 2D effective theory technique to compute the partition function at  low temperatures. An  interesting result from this approach is the emergence of a quantum correction of the form $\log{T}$ in the logarithm of partition function. It has been shown that, once this correction is taken into account, the average i.e. thermodynamic energy remains greater than that of Hawking radiation even at small temperatures. Hence, it was concluded that there is actually no mass gap. In a nutshell, due to the breakdown of semiclassical analysis at low enough temperatures, it is required to consider the effect of quantum corrections. In this section, we shall address the same in the original 4D description of the near-extremal black holes.

\subsection{Form of the quantum corrections in near-extremal limit}

We attempt to understand the one-loop correction to the partition function for a near-extremal black hole via a Euclidean path integral computation in 4D, without getting into an effective lower-dimensional description. The near-extremal solution is parametrized by two large parameters: the charge (or extremal horizon radius) $r_0$ and the inverse temperature $\beta \sim 1/T$. We evaluate the large contributions involving these parameters, in particular, the logarithmic contributions. Although computing the full one-loop contribution directly is out of hand, Sen and collaborators have put forward \cite{Banerjee:2010qc, Banerjee:2011jp, Sen:2012kpz, Sen:2012cj, Bhattacharyya:2012wz, Sen:2012dw} a general strategy to extract the logarithm of horizon radius contributions for (non-)extremal black holes. As we will argue below, the $\log{T}$ contributions cannot be obtained by taking a small temperature limit of these computations. Toward the end of this section, we present our strategy to compute such corrections. We find that, to the leading order, the large quantum contributions are of the form $\log{r_0}$ and $\log{T}$, whereas there are further polynomially suppressed corrections in temperature.

\subsection{A brief discussion on the log correction for (non-)extremal black holes}

Following \cite{Sen:2012kpz, Bhattacharyya:2012wz}, to compute the one-loop partition function for a generic black hole solution in Einstein-Maxwell theory \eqref{EM-act}, the fields are fluctuated around the black hole background, 
\begin{align}
    g_{AB} = \tilde{g}_{AB} + h_{AB}, \quad A_{B} = \tilde{A}_B + \frac{1}{2}a_B .
\end{align}
The action is expanded to quadratic order in fluctuations. The zeroth order term of the expansion is the on-shell action, evaluated for the background  $\{\tilde{g}_{AB},\tilde{A}_{B}\}$, which is a constant and needs to be regulated properly to get sensible semiclassical physics. By action principle, in the presence of appropriate boundary terms \eqref{EM-bdy}, the first-order term vanishes as the background satisfies the equations of motion. Our goal is to integrate out the Gaussian-like quadratic action and find the one-loop correction to the partition function. 

Since the fluctuations have redundancies due to diffeomorphism and $U(1)$ gauge invariances, we also add gauge-fixing terms of the following form, to the quadratic action,
\begin{align}
    S_{\text{diffeo}} &= -\frac{1}{2}\int d^4x \sqrt{\tilde{g}}\left(\tilde{\nabla}_A h^{AC} - \frac{1}{2}\tilde{\nabla}^C h \right)\left(\tilde{\nabla}^B h_{BC} - \frac{1}{2}\tilde{\nabla}_C h \right), \\
    S_{\text{gauge}} &= -\frac{1}{2}\int d^4x \sqrt{\tilde{g}} (\tilde{\nabla}_A a^A)^2.
\end{align}
The quadratic action of fluctuations takes the form,
\begin{align}\label{TMA}
    S^{(2)} \equiv \int d^4x \sqrt{\tilde g}\ \Psi \tilde{\Delta} \Psi,
\end{align}
where $\Psi$ represents all the fields of the theory and $\tilde{\Delta}$ is a 2-derivative differential operator, constructed out of the background. The partition function is then given as the integral,
\begin{align}
    Z = \int \mathcal{D}\Psi\text{e}^{-S^{(2)}} = \frac{1}{\sqrt{\text{det}(\tilde{\Delta})}}.
\end{align}
We have omitted the constant semiclassical contribution to avoid notational clutter. To evaluate the integral it is required to compute the eigenvalues of the kinetic operator which in turn gives the determinant. Using the heat-kernel formalism for a generic (non-) extremal background, presented in \cite{Bhattacharyya:2012wz, Sen:2012dw}, the logarithm of horizon radius contribution can be computed. In principle, for the computation of partition function, the Lagrangian density should be integrated over the full background. Due to the infinite AdS$_2$ throat in the near-horizon region of an extremal black hole, the dynamics is wonderfully captured by the near-horizon geometry. Hence, for an extremal black hole, the background is considered to be the near horizon AdS$_2\times$S$^2$ geometry. An important point to note is that, for non-extremal black holes, one needs to remove the effects of thermal gas to obtain the correct entropy corresponding to the degeneracy of the black hole states. 

For an extremal black hole, the log correction can be computed even without the heat-kernel method. Since for the extremal AdS$_2\times$S$^2$ background, the eigenfunctions of the kinetic operator are known. Using the explicit form of these eigenfunctions, the log correction has also been computed by finding the eigenvalues for a class of extremal black holes \cite{Banerjee:2010qc, Banerjee:2011jp, Sen:2012kpz, Sen:2012cj}. These corrections are also computed using Sen's quantum entropy function formalism\cite{Sen:2008yk, Gupta:2008ki, Banerjee:2010qc}.

For a near-extremal black hole, it is natural to consider a small temperature limit of the non-extremal result. The computation for a non-extremal black hole \cite{Sen:2012dw} is however performed under a limit where the horizon radius $r_{+}$ and the inverse temperature $\beta$ are of the same order i.e. $r_{+}\sim\beta$. This is not true for a near-extremal black hole, where the full horizon radius depends on two independent large parameters: the extremal radius and inverse temperature. Also, this computation gives the temperature-dependent corrections to be a polynomial expansion. Through this procedure, it is not possible to obtain the $\log{T}$ corrections. Therefore, we consider the near-extremal black hole as a deviation from the extremal one and try to compute the $\log{T}$ corrections. We discuss our strategy for the same in the next subsection.

\subsection {Strategy for the quantum correction computation for near-extremal black holes} \label{pert-theory-sec}
We compute the one-loop corrected partition function for a near-extremal black hole by finding the eigenvalues of the kinetic operator. We consider the near-horizon region of the black hole to be a small temperature deviation of the extremal near-horizon geometry. The near-horizon throat of an extremal black hole is infinite and hence, all the computations for an extremal black hole get contributions from the near-horizon region only. For a near-extremal black hole, the throat is finite yet large. Therefore, we expect that many of the physical questions can be answered from the near-horizon region. In the far region near asymptotic infinity, the geometry can be well-approximated by the full extremal geometry. Also, the fluctuations die off in this region. Therefore, in presence of the large near-horizon throat, the contributions coming from the FHR are very small compared to the contributions of the NHR. Hence, we focus on the near-horizon physics, where the near-extremal geometry is a perturbative linear order temperature deviation of AdS$_2\times$S$^2$ geometry and is given in \eqref{NE-corr}. The kinetic operator can also be expanded in the same way. This allows us to apply first-order perturbation theory for the computation of the eigenvalues. The computation is schematically described below. 

Due to the perturbative expansion of the background geometry, the kinetic operator splits into two parts given as $\tilde{\Delta} = \Delta^0 +T\Delta^{(c)}$. The $\mathcal{O}(T^0)$ term $\Delta^0$ is the extremal kinetic operator. Whereas the $\mathcal{O}(T)$ term $\Delta^{(c)}$ is a differential operator which we treat perturbatively. We denote the eigenvalues of the full kinetic operator by $\Lambda_n$, which are small deviations from the eigenvalues of the extremal operator as,
\begin{align}
    \tilde{\Lambda}_n = \Lambda^0_n + T\Lambda^{(c)}_n.
\end{align}
Here $\Lambda^0_n$ are the eigenvalues of the extremal kinetic operator such that,
\begin{align}
    & \Delta^0 f^0_n(x) = \Lambda^0_n f^0_n (x) ,
\end{align}
where, $f^0_n(x)$ represents the orthonormal eigenfunctions of the operator $\Delta^0$. Now we invoke the standard machinery of first-order perturbation theory. We start with the modified eigenvalue equation having the following form,
\begin{align}
    (\Delta^0 + T \Delta^{(c)}) (f^0_n(x) + T f_n^{(c)}(x)) = (\Lambda^0_n + T \Lambda^{(c)}_n)(f^0_n (x) + T f_n^{(c)} (x)). \label{pert-eigen-eqn}
\end{align}
The $\mathcal{O}(1)$ terms vanish due to the eigenvalue equation of the extremal kinetic operator. Thus at $\mathcal{O}(T)$, we have:
\begin{align}
    \Delta^{(c)} f^0_n + \Delta^0 f_n^{(c)} = \Lambda^{(c)}_n f^0_n + \Lambda^0_n f_n^{(c)}.
\end{align}
Taking inner product with $f_n^{0*}$ on both sides of the equation and using the orthonormality conditions we have the correction to the eigenvalues as,
\begin{align}
    \Lambda^{(c)}_n = \int d^4x \sqrt{g^0}\ f^{0*}_n(x)\ \Delta^{(c)}\ f^0_n(x).
\end{align}
In order to find the corrections to eigenfunctions, we take inner product of \eqref{pert-eigen-eqn} with $f_m^{0*}$ for $m\neq n$, which gives the following correction,
\begin{align}
    f_n^{(c)} (x) = \sum_{m\neq n}\frac{1}{\Lambda^0_n - \Lambda^0_m}\left(\int d^4x' \sqrt{g^0}\ f^{0*}_m(x')\ \Delta^{(c)}\ f^0_n(x')\right)\ f_m^0 (x).
\end{align}

To find the one-loop determinant, only the evaluation of the eigenvalues is required. The one-loop correction to the logarithm of partition function can be computed for $\tilde{\Lambda}_n \neq 0$ as given by,
\begin{align}
     \log{Z} = - \frac{1}{2}\sum_n \log{\tilde{\Lambda}_n}. 
\end{align}

\subsubsection*{Contribution from extremal zero modes:}
We consider the eigenfunctions of the extremal kinetic operator, which have zero eigenvalues i.e. $\Lambda^0_n = 0$. For these modes, the corrected eigenvalues are linear in temperature. Therefore, the extremal zero modes acquire some small non-zero mass in the near-extremal background. These modes contribute to the $\log{T}$ corrections in the logarithm of the partition function.

\subsubsection*{Contribution from extremal non-zero modes:}
From the non-zero modes of the extremal kinetic operator, we will get contributions of the form $\log{r_0} +  \mathcal{O}(T)$ in the expression of the logarithm of the partition function. These corrections are much suppressed as compared to the $\log{T}$ contribution.

\subsubsection*{Contribution from near-extremal zero modes:}
There might be some modes that are zero modes for both extremal and near-extremal backgrounds. For such modes, the eigenvalue correction is $\mathcal{O}(T^2)$. Because of the vanishing eigenvalues, we cannot perform the corresponding Gaussian integrals. These modes can affect the partition function only through the measure. We will impose normalization conditions on these zero modes similar to the standard prescription, and investigate the contributions. As we will see later, there are indeed these zero modes but their measure does not give $\log{T}$ contribution.

From this analysis, we understand that the $\log{T}$ correction should be given by the contributions of the modes which are exact zero modes of the extremal kinetic operator. The origin of this correction is the small temperature-dependent mass acquired by the zero modes in presence of near-extremal correction to the background geometry. In the next section, we undertake this approach.

\section{Computation of $\log{T}$ contributions} \label{NE-logT-sec}

In this section, we will compute the eigenvalues for the kinetic operator on the near-horizon near-extremal background using first-order perturbation theory and find the $\log{T}$ corrections. Firstly, we consider the quadratic action \cite{Bhattacharyya:2012wz, Sen:2012kpz} for the fluctuations $\{h_{AB}, a_{A}\}$. Quadratic Lagrangian density for graviton,
\begin{align}
    \mathcal{L}_{hh} &= h_{AB} \Big[\frac{1}{4}\tilde g^{AC}\tilde g^{BD}\tilde \Box - \frac{1}{8}\tilde g^{AB}\tilde g^{CD}\tilde \Box + \frac{1}{2}\tilde R^{ACBD} + \frac{1}{2}\tilde R^{AC}\tilde g^{BD} - \frac{1}{2}\tilde R^{AB}\tilde g^{CD} \nonumber \\
    & +\frac{1}{8}\tilde F^2\left( 2\tilde g^{AC}\tilde g^{BD} - \tilde g^{AB}\tilde g^{CD} \right)  - \tilde F^{AC}\tilde F^{BD} - 2\tilde F^{AE} \tensor{\tilde F}{^C_E}\tilde g^{BD} + \tilde F^{AE}\tensor{\tilde F}{^B_E}\tilde g^{CD} \Big] h_{CD}.
\end{align}
Quadratic Lagrangian density for photon,
\begin{align}
    \mathcal{L}_{aa} = \frac{1}{2}a_A\left(\tilde g^{AB}\tilde \Box - \tilde R^{AB}\right)a_B.
\end{align}
Mixing terms between graviton and photon,
\begin{align}
    \mathcal{L}_{ha} = -h_{AB}\left( 4\tilde g^{A[C}\tilde F^{D]B} + \tilde F^{CD}\tilde g^{AB} \right)\tilde \nabla_C a_D.
\end{align}
Ghost Lagrangian,
\begin{align}
    \mathcal{L}_{\text{ghost}} = b_A\left(\tilde g^{AB}\tilde \Box + \tilde R^{AB}\right)c_B + b\tilde \Box c - 2b\tilde F^{AB}\tilde \nabla_A c_B.
\end{align}
We have added the ghost terms to the action due to gauge fixing. Here the background is taken to be near-extremal. Therefore, the full quadratic action is given as, 
\begin{align}
    S = \int d^4x \sqrt{\tilde g}(\mathcal{L}_{hh}+\mathcal{L}_{aa}+\mathcal{L}_{ha}+\mathcal{L}_{\text{ghost}}). \label{ext-quad-act}
\end{align}

\subsection{The extremal zero modes}

For the quantum correction to the partition function, we need to find all the corrected eigenvalues. As discussed earlier, the zero modes of extremal background can give rise to $\log{T}$ correction, whereas the nonzero modes give rise to polynomial corrections suppressed by powers of $T$. In the appendix \ref{basis-app}, we have reviewed the eigenfunctions of the extremal kinetic operator. There are two classes of normalizable eigenfunctions on AdS$_2$ which are labeled by some continuous and discrete parameters. The discrete modes physically correspond to large gauge transformations and large diffeomorphisms, whereas the continuous modes are derived from normalizable scalars. Although the large gauge transformations and large diffeomorphisms are non-normalizable, the discrete vector and tensor modes, constructed out of their derivatives, are normalizable. The zero modes are part of the discrete modes \cite{Sen:2012kpz}. See also \cite{Iliesiu:2022onk}, for a detailed discussion on the zero modes and their regularization. 

Because of orthogonality, all the modes decouple in the extremal background hence their contributions can be studied separately. Firstly, we consider the contributions from discrete modes and identify the zero modes amongst them. We expand the nonzero components of the fields following \cite{Sen:2012kpz} as linear combinations of discrete eigenfunctions,
\begin{align}
    & a_\mu = E_1 v_\mu + E_2 \varepsilon_{\mu\nu}v^\nu, \nonumber \\
    & h_{\mu i} = \frac{1}{\sqrt{\kappa}}\left(E_3\partial_i v_\mu + \tilde{E}_3\varepsilon_{\mu\nu}\partial_i v^\nu + E_4\varepsilon_{ij}\partial^j v_\mu + \tilde{E}_4\varepsilon_{ij}\varepsilon_{\mu\nu}\partial^j v^\nu\right), \nonumber \\
    & h_{\mu\nu} = \frac{r_0}{\sqrt{2}}(\nabla_\mu\hat{\xi}_\nu + \nabla_\nu\hat{\xi}_\mu - g_{\mu\nu}\nabla^\rho\hat{\xi}_\rho) + E_6 w_{\mu\nu}; \quad \hat{\xi}_\mu = E_5 v_\mu + \tilde{E}_5 \varepsilon_{\mu\nu}v^\nu.
\end{align}
Here, $v_\mu$ is the normalizable vector mode \eqref{non-norm} constructed out of the discrete non-normalizable scalar modes, multiplied with spherical harmonics. $w_{\mu\nu}$ is the discrete normalizable tensor mode \eqref{non-norm_tensor} corresponding to non-normalizable diffeomorphisms, multiplied with the spherical harmonics. $\kappa$ is the $-\Box_{S^2}$ eigenvalue given as $\frac{l(l+1)}{r_0^2}$. We have suppressed the mode labels for simplicity since the different labels do not mix among themselves. For each sector, we will evaluate the contribution to the action, and finally, we will take a sum over all modes. 

In the $l=0$ sector of spherical harmonics, the modes $E_3, \tilde{E}_3, E_4, \tilde{E}_4$ are absent since these modes involve derivatives on $S^2$. Therefore, the contribution to the zeroth order (i.e. extremal) action is given as,
\begin{align}
    -\frac{1}{2}\kappa (E_1^2 + E_2^2) -\frac{1}{2}(\kappa + 2r_0^{-2})(E_5^2 + \tilde{E}_5^2) -\frac{1}{2}\kappa E_6^2. 
\end{align}
The contribution is diagonal in the coefficients $E_i$ i.e. the corresponding basis elements are eigenfunctions of the extremal kinetic operator. Since $\kappa = 0$ for $l=0$, we see that the contributions coming from $E_1, E_2, E_6$ vanish. Hence, the corresponding basis elements i.e. $v_\mu, \varepsilon_{\mu\nu}v^\nu, w_{\mu\nu}$ respectively, are the zero modes of the extremal operator. We will find the correction of eigenvalues for these eigenfunctions.

The contribution to the zeroth order action coming from each sector corresponding to $l\geq 1$ is given as,
\begin{align}
    & -\frac{1}{2}\kappa (E_1^2 + E_2^2) -\frac{1}{2}(\kappa + 2r_0^{-2})(E_5^2 + \tilde{E}_5^2) -\frac{1}{2}\kappa E_6^2 \nonumber \\
    &  -\frac{1}{2}(\kappa - 2r_0^{-2})(E_3^2 + \tilde{E}_3^2 + E_4^2 + \tilde{E}_4^2) + 2i r_0^{-1}\sqrt{\kappa}(E_1\tilde{E}_3 - E_2 E_3).
\end{align}
The modes corresponding to $E_1, \Tilde{E}_3$ and $E_2, E_3$ mix amongst themselves. For $l=1$, the $E_4, \Tilde{E}_4$ terms vanish i.e. the corresponding basis elements are zero modes of the extremal operator. Beyond $l=1$, all modes have nonzero eigenvalues.

\subsubsection{Contribution from $l = 0$ tensor modes} \label{tensor-sec}
The tensor modes $w^n_{\mu\nu}$ are degenerate in the discrete label $n$. Therefore, we apply degenerate perturbation theory to find the matrix elements between different labels. This matrix turns out to be diagonal. The eigenvalue correction corresponding to $w^n_{\mu\nu}$ is given by the integral of $w^n\cdot\Delta\cdot w^n$:
\begin{align}
    \Lambda[w^n_{\mu\nu}] = \frac{n\pi T}{256r_0}\Big[&-69 + 8n(-6+11n+8n^2) + 4(1+n)(-1+8n^2)\cosh{\eta_0} + \nonumber \\
    & + 4(1+4n+2n^2)\cosh{2\eta_0} + 4(1+n)\cosh{3\eta_0} + \cosh{4\eta_0} \Big]\nonumber\\
    &\cdot\left(\sech{\frac{\eta_0}{2}}\right)^6\left(\csch{\frac{\eta_0}{2}}\right)^2(\coth{\eta_0}+\csch{\eta_0})^{-2n}.
\end{align}
Using the value of the radial cutoff $\eta_0$ from \eqref{NH-rad-cutoff}, we get,
\begin{align}
    \Lambda[w^n_{\mu\nu}] = \frac{n\pi T}{2r_0}. 
\end{align}
This is the first-order correction to the eigenvalue for the tensor modes. The contribution to the logarithm of the partition function, coming from the tensor zero modes\footnote{The real and imaginary parts of the tensor modes have the same eigenvalues. Hence, we multiply with a factor of 2.} is given as,
\begin{align}
    \log{Z}_{\text{tensor}} &= - 2\cdot\frac{1}{2}\sum_{n\geq 2} \log{\Lambda[w^n_{\mu\nu}]} \nonumber \\
    & = - \sum_{n\geq 2} \log{\left(\frac{n\pi T}{2r_0}\right)} \nonumber\\
    & = \log{\left(\prod_{n\geq 2} \frac{2r_0}{n \pi T}\right)}.
\end{align}
The product over $n$ inside the logarithm can be evaluated using zeta function regularization \cite{Saad:2019lba, Moitra:2021uiv},
\begin{align}
    \prod_{n\geq 2} \frac{\alpha}{n T} = \frac{1}{\sqrt{2\pi}}\ \frac{T^{3/2}}{\alpha^{3/2}}. \label{prod}
\end{align}
Using this result to compute the product, we have:
\begin{align}
    \log{Z}_{\text{tensor}} \sim \frac{3}{2}\log{T}. \label{tensor-corr}
\end{align}
The contribution coming from tensor zero modes agrees with the effective 2D theory results as derived in \cite{Iliesiu:2020qvm, Iliesiu:2022onk}.
 The contributions to the partition function due to the modified eigenvalues of the extremal tensor zero modes can also be derived from the exact quantization of a Schwarzian theory. We come back to this discussion in section \ref{1D-sec}. The reason behind getting the same contribution from a one-loop computation stems from the one-loop exact structure of the Schwarzian theory. But the one-loop action \eqref{TMA} for the orthonormalized tensor modes does not reproduce the Schwarzian action. The emergence of a Schwarzian-like action from the tensor zero modes has been discussed in \cite{Iliesiu:2022onk} where the authors have used a particular normalization for the modes. It differs from that of the standard orthonormal basis discussed in \cite{Sen:2012kpz}, which we have used extensively for our work. The computation of the  action that describes the tensor zero modes requires an effective description of the theory, as will be described in section \ref{1D-sec}. 

\subsubsection{Contribution from $l = 0$ vector modes}
We denote the vector modes as, $v^{a,n}_{\mu}\equiv\{v^n_\mu, \varepsilon_{\mu\nu}v^{n,\nu}\}$, where $n$ is the discrete label. All these modes are degenerate, therefore we invoke degenerate first-order perturbation theory. Hence we find the matrix elements:
\begin{align*}
    \int d^4x \sqrt{g}\  v^{a,p} \cdot \Delta \cdot v^{b,n}, 
\end{align*}
here $\Delta$ is the kinetic operator, with an appropriate spacetime index structure. It turns out that this matrix is diagonal i.e. proportional to $\delta^{pn}\delta_{ab}$. For the eigenvector $v^n_\mu$, we find the eigenvalue:
\begin{align}
    \Lambda[v^n_\mu] = \frac{n\pi T}{2r_0}(1+2n+n\cosh{\eta_0})\left(\sech{\frac{\eta_0}{2}}\right)^4 \left(\tanh{\frac{\eta_0}{2}}\right)^{2n}.
\end{align}

The eigenvalue corresponding to the eigenvector $\varepsilon_{\mu\nu}v^{n,\nu}$ is given as, ${\Lambda[\varepsilon_{\mu\nu}v^{n,\nu}] = \Lambda[v^n_\mu]}$. Using the value of the radial cutoff $\eta_0$ in \eqref{NH-rad-cutoff}, at first order in temperature, the eigenvalue is 0 since $\Lambda[v^n_\mu] \sim \mathcal{O}(T^2)$. Therefore, we conclude that these modes are zero modes even in the near-extremal background and we cannot perform a Gaussian integral over them.

To understand the structure of the contribution to the partition function coming from the measure of these zero modes, we consider the normalization condition,
\begin{align}
    \int \mathcal{D}a_{\mu}\text{exp}\left(-\int d^4x \sqrt{g}g^{\mu\nu}a_{\mu}a_{\nu} \right) = 1.
\end{align}
Here we have considered the fluctuations $a_{\mu}$ to be a linear combination of the $l = 0$ vector zero modes given as $a_{\mu} = \alpha_n v^n_{\mu}$. Since these modes are also zero modes of the extremal background, we can readily see that the exponent in this integration has a temperature-independent piece and a term, linear in temperature. We get this form using the orthogonality condition of the modes. Considering $\mathcal{D}a_{\mu}\sim \mathcal{N}'\prod_n d\alpha_n$, the normalization condition has the following form,
\begin{align}
    & \int \mathcal{N}'\prod_n d\alpha_n \text{exp}(-\mathcal{N}_n^2\alpha_n^2) = 1. 
\end{align}
Performing the Gaussian integral, we have
\begin{align}
\frac{\mathcal{N}'}{\sqrt{\prod_n\mathcal{N}_n^2}} = 1, \quad \mathcal{N}' = \prod_n\mathcal{N}_n \sim \mathcal{O}(1) + \mathcal{O}(T).
\end{align}
Therefore, we get that the form of the contribution coming from the measure has a $\mathcal{O}(1)$ i.e. a temperature independent piece. In other words, there is no factor of $T$ multiplying the partition function, hence giving no $\log{T}$ contribution to the logarithm of partition function. These contributions will be polynomially suppressed in temperature.

\subsubsection{Contribution from $l = 1$ vector modes}
We denote these modes as $y^{a,n}_{\mu i} = v^{a,n}_{\mu}\xi^{2;1,m}_{i} \equiv \{\frac{1}{\sqrt{\kappa}}\varepsilon_{ij}\partial^j v_\mu , \frac{1}{\sqrt{\kappa}}\varepsilon_{ij}\varepsilon_{\mu\nu}\partial^j v^\nu\}$. Here $\kappa = 2r_0^{-2}$ is the $-\Box_{S^2}$ eigenvalue for the $l=1$ sector and $\xi^{2;1,m}_{i}$ is a vector eigenfunction of the Laplacian on $S^2$ as in \eqref{vector_basis_S2}. Clearly, $m$ runs over the values $-1, 0, +1$. Again we invoke degenerate perturbation theory but the correction matrix turns out to be diagonal. Therefore, for each value of the labels $\abs{m}\leq 1$ and $ n\geq 1$, we have the correction corresponding to $\varepsilon_{ij}\partial^j v_\mu$:
\begin{align}
    \Lambda[\varepsilon_{ij}\partial^j v_{\mu,n}] = \frac{n\pi T}{32r_0}[7+8n+4(1+n)\cosh{\eta_0}+\cosh{2\eta_0}]\left(\sech{\frac{\eta_0}{2}}\right)^4 \left(\tanh{\frac{\eta_0}{2}}\right)^{2n}.
\end{align}
The eigenvalue correction corresponding to the second kind of eigenfunction is the same i.e. and to order $T$, the value is given by,
\begin{align}
    \Lambda[\varepsilon_{\mu\nu}\varepsilon_{ij}\partial^j v_n^\nu] = \Lambda[\varepsilon_{ij}\partial^j v_{\mu,n}] =  \frac{n\pi T}{4r_0}.
\end{align}
The contributions from these modes to the partition function are given by,
\begin{align}
    \log{Z}_{l = 1 \, \text{vector}} &= - \frac{1}{2}\sum_{\substack{n\geq 1, \\ \abs{m} = 0,1}}\log{\Lambda[\varepsilon_{ij}\partial^j v_{\mu,n,m}]} - \frac{1}{2}\sum_{\substack{n\geq 1, \\ \abs{m} = 0,1}}\log{\Lambda[\varepsilon_{\mu\nu}\varepsilon_{ij}\partial^j v_{n,m}^\nu]} \nonumber \\
    & = -\frac{6}{2}\sum_{n\geq 1} \log{\left(\frac{n\pi T}{4r_0}\right)} \nonumber \\
    & = 3\log{\left(\prod_{n\geq 1} \frac{4r_0}{n \pi T}\right)}.
\end{align}
Using \eqref{prod}, we compute the product inside the logarithm, where we consider the $n = 1$ contribution separately. Therefore, we have
\begin{align}
    \log{Z}_{l = 1 \, \text{vector}} = 3\log{\left(\frac{\pi^{3/2}}{\sqrt{2\pi}}\frac{T^{3/2}}{(4r_0)^{3/2}}\right)} + 3\log{\left(\frac{4r_0}{\pi T}\right)}.
\end{align}
Therefore, we also have $\log{T}$ contribution from the $l = 1$ zero modes, given by:
\begin{align}
    \log{Z}_{l = 1 \, \text{vector}} \sim \frac{3}{2}\log{T}. \label{vector-corr}
\end{align}

\subsection{Total $\log{T}$ contribution from extremal zero modes}
From our analysis, we get that the tensor modes give rise to the $\log{T}$ contribution that matches with the Schwarzian result. The $l = 0$ vector modes have zero contribution at first-order in temperature. Whereas, the $l = 1$ vector modes give a non-trivial contribution. The full contribution is given by,
\begin{align}
    \log{Z} = \log{\left(\frac{\pi^{3/2}}{\sqrt{2\pi}}\frac{T^{3/2}}{(2r_0)^{3/2}}\right)} + 3\log{\left(\frac{\pi^{3/2}}{\sqrt{2\pi}}\frac{T^{3/2}}{(4r_0)^{3/2}}\right)} + 3\log{\left(\frac{4r_0}{\pi T}\right)}.
\end{align}
Hence, the dependence from \eqref{tensor-corr} and \eqref{vector-corr} is given as,
\begin{align}
    \log{Z}\sim 3\log{T}. \label{full-logT}
\end{align}
The corrections coming from all other modes at first-order in temperature are suppressed. The large contribution coming from the charge of the black hole can be found in \cite{Sen:2012kpz}.

\section{Revisiting the 1D effective description} \label{1D-sec}
In this section, we revisit the computation of the $\log{T}$ corrections to the logarithm of partition function from an effective theory description. In particular, we  show that the physics of the tensor zero modes at low temperatures is described by a  Schwarzian theory. For addressing this description, working in the s-wave sector of the fields would be enough. We first reduce the theory \eqref{EM-act} along with the boundary terms \eqref{EM-bdy} located at the asymptotic boundary of a spherically symmetric Euclidean black hole. In order to understand the quantization of the system, we follow the decomposition of the near-extremal geometry into near-horizon and far-horizon regions as in section \ref{NHR-sec}. Because of the long near-horizon throat, the quantum fluctuations in the FHR are suppressed as compared to the fluctuations in the NHR. Hence, we put the action on-shell in FHR and this effectively \textit{induces} a local boundary term at the boundary separating the NHR and FHR, as discussed in the appendix \ref{RN-thermo-app}. 
To understand the quantization at the NHR region we take the following strategy: 

\begin{itemize} 

\item {\bf Finding the 2D effective action: \,\,}Since our interest is in spherically symmetric near-extremal black holes, we first reduce the 4-dimensional Einstein-Hilbert theory on an arbitrary spherically symmetric background.
This gives us a reduced theory on a 2D manifold. Working in the s-wave sector, we consider the dimensional reduction ansatz as:
\begin{align}
    ds^2 = \frac{r_0}{\Phi}g_{\mu\nu}dx^\mu dx^\nu + \Phi^2(x)(d\psi^2+\sin^2{\psi}d\varphi^2), \quad A_B \equiv (A_{\mu},0). \label{dim-red-ansatz}
\end{align}
Plugging this ansatz into the action, we get a 2D Einstein-Hilbert-Maxwell action non-minimally coupled to the scalar $\Phi$. The Weyl factor of the 2D metric is so chosen that the kinetic term of the scalar vanishes. Integrating out the 2D gauge fields, we obtain the 2D effective theory,
\begin{align}
    \mathcal{S} = -4\pi\int_N d^2x \sqrt{g}\left(\Phi^2 R + \frac{2r_0}{\Phi}-\frac{2r_0^3}{\Phi^3}\right) - 8\pi\int_{\partial N} dx \sqrt{\gamma}\Phi^2 K. \label{2Dact}
\end{align}

The variational problem is well-defined for this theory when we impose Dirichlet boundary conditions on the fields. It admits a classical solution  given by an AdS$_2$ metric and a constant dilaton as,
\begin{align}
    g_{\mu\nu}dx^\mu dx^\nu = r_0^2(d\eta^2+\sinh^2{\eta}d\theta^2), \quad \Phi =  r_0. \label{ext-2D}
\end{align}
This solution can be uplifted to the 4D extremal near-horizon geometry \eqref{ext-NH}.
\item 
{\bf Finding the near-extremal background:\,\,} Next, we look for another classical solution of this theory, which is a deviation from the solution \eqref{ext-2D} by a small temperature. We demand that, once obtained, the same  should be uplifted to the near-horizon geometry of a near-extremal black hole in the four-dimensional parent theory. To get the same, first, we consider a deviation from extremality \eqref{ext-2D} as,
\begin{align}
    \bar{g}_{\mu\nu}dx^\mu dx^\nu = r_0^2(d\eta^2+\sinh^2{\eta}d\theta^2) + \delta g, \, \Phi = r_0(1+\phi),
\end{align}
such that the variations $\delta g$ and $\phi$ do not die off at the boundary $\partial N$. Expanding the action \eqref{2Dact} in these deviations and solving the equations of motion corresponding to these fields $\delta g$ and $\phi$, we intend to find the background solution that uplifts to the near-horizon near-extremal background as given in \eqref{NE-corr}. The expansion of the action is given as,
\begin{align}
    \mathcal{S} = 16\pi^2 r_0^2 - 16\pi \int_{\partial N} \sqrt{\gamma}\phi K + \mathcal{S}^{(2)}[\delta g,\phi] .\label{act-fluc}
\end{align}
The second-order action $\mathcal{S}^{(2)}$ is important to understand the structure of $\delta g_{\mu\nu}\equiv\sigma_{\mu\nu}$ and $\phi$ by solving the equations of motion for which only the bulk action is enough. The bulk part of the same is given below,
\begin{align}
    \mathcal{S}^{(2)}_{\text{bulk}} = \int d^2x & \sqrt{g}r_0^2 \Big[\frac{1}{4r_0^2}\sigma^2 - \frac{1}{2r_0^2}\sigma_{\mu\nu}\sigma^{\mu\nu} +\frac{1}{2}\sigma\nabla_\mu\nabla_\nu\sigma^{\mu\nu} - \frac{1}{4}\sigma\nabla^2\sigma + \frac{1}{4}\sigma^{\mu\nu}\nabla^2\sigma_{\mu\nu} \nonumber\\
    & - \frac{1}{2}\sigma^{\nu\rho}\nabla_\mu\nabla_\nu\sigma^{\mu}_{\rho} + 2\phi (\nabla_\mu\nabla_\nu\sigma^{\mu\nu}-\nabla^2\sigma + \frac{1}{r_0^2}\sigma)  -\frac{12}{r_0^2}\phi^2 \Big]. \label{2d-quad-act}
\end{align}

Here we note that at the first-order in variation, the action is a pure boundary term depending only on the dilaton variation $\phi$ and it is constant on the boundary. Furthermore, even though $\delta g$ does not vanish at the boundary, all other first-order terms depending on $\delta g$ vanish\footnote{ This is a consequence of the simple structure of 1D boundary for which the extrinsic curvature is a pure trace i.e. in terms of boundary coordinates, $K_{ab} = K\gamma_{ab}$. }.

Now we turn to find the near-extremal solution such that the deviation from extremality correctly uplifts to \eqref{NE-corr}. To get that, the arbitrary deviations $\delta g$ may be decomposed into pure trace and traceless parts \cite{DHOKER1986205, Moitra:2021uiv}, where the trace is computed with respect to the AdS$_2$ metric \eqref{ext-2D}. Comparing \eqref{NE-corr} and the ansatz \eqref{dim-red-ansatz}, we notice that for the near-extremal solution, the deviation of the 2D metric (i.e. $\frac{r_0}{\Phi}\bar{g} - g$) should be traceless. This fixes the trace of $\bar{g}$ in terms of the dilaton field. Maintaining these,  we consider the form of the deviation as,
\begin{align}
    \delta g_{\mu \nu}\ dx^{\mu}dx^{\nu} = \phi(\eta)(d\eta^2 + \sinh^2{\eta}d\theta^2) + \alpha(\eta)(d\eta^2 - \sinh^2{\eta}d\theta^2) .
\end{align}
Here we have taken a static ansatz i.e. the corrections are independent of $\theta$. The equations of motion coming from the second-order action \eqref{2d-quad-act} are,
\begin{align}
    & \tanh{\eta}\ \phi'' - \phi' = 0, \nonumber \\
    & \alpha'' + 3\coth{\eta}\ \alpha' + \alpha = 4\phi'' + 4(3r_0^2 - 1)\phi .\label{2Deom}
\end{align}
Choosing appropriate integration constants and taking care of the Weyl factor, it can be shown that a generic solution of these equations gets uplifted to the solution described in \eqref{NE-corr} with the functions  $\alpha, \phi$ given as,
\begin{align}
    & \phi = 4\pi r_0^3 T \cosh{\eta} , \quad
     \alpha = 4\pi r_0^3 T (2 +\cosh{\eta})\tanh^2{\left(\frac{\eta}{2}\right)} .
\end{align} \nonumber
\item{\bf Quantization of the linear order action: \,\,}
Finally to quantize the theory at one-loop order around the above background, we consider the first-order deviation term of the action. The boundary behavior of the dilaton $\phi$ can be fixed from the near-extremal solution. The presence of near-extremal deviations makes the asymptotic symmetry modes of AdS$_2$ slightly nondegenerate. These modes can be realized as a nontrivial wiggly-shaped boundary on rigid AdS$_2$ and the shape of the boundary can be parametrized by an arbitrary function $\theta(u)$, where $u$ is the boundary coordinate. The linear-order boundary term in \eqref{act-fluc} corresponds to the effective action of these boundary gravitons. It is well-studied in the literature that this boundary theory gives rise to a Schwarzian action \cite{Maldacena:2016upp, Saad:2019lba} of boundary modes\footnote{See also \cite{Banerjee:2021vjy} for a review on this boundary description.}. This action has the form $\int du\  \text{Sch}\left(\tan{\frac{\theta}{2}},u\right)$, where the Lagrangian density is a Schwarzian derivative\footnote{The Schwarzian derivative is defined as, $$\text{Sch}(F,u) = - \frac{1}{2}\left(\frac{F''}{F'}\right)^2 + \left(\frac{F''}{F'}\right)'.$$}. The theory is also one-loop exact \cite{Stanford:2017thb}, which allows us to compute the partition function exactly when we consider the leading order deviation from extremality \cite{Iliesiu:2020qvm}. The contribution to the logarithm of the partition function turns out to be,
\begin{align}
    \log{Z}\sim \frac{3}{2}\log{T}.
\end{align}
This contribution can be traced back to the tensor zero modes contribution discussed in \eqref{tensor-corr}.
The density of states \cite{Stanford:2017thb, Saad:2019lba} from this computation gives a dependence of $\sinh{2\sqrt{E}}$ and it smoothly vanishes to zero as $E\rightarrow 0$.
This effective description does not incorporate the polynomially suppressed contributions in temperature to the logarithm of the partition function.

\end{itemize}

Thus we find that the quantum (tensor modes) corrections to the partition function of near-extremal black holes can be computed from a direct four-dimensional analysis as in section \ref{tensor-sec} and from an effective two-dimensional analysis as in section \ref{1D-sec}.
We would like to emphasize some points while comparing these two descriptions. To get an effective description, we fluctuate the fields around the extremal background, where the fluctuations do not die on the boundary. To get the correct near-extremal geometry, we consider the second-order action and solve the equations of motion. The analysis also shows us that the near-horizon geometry of the near-extremal black hole is not locally AdS$_2$. In fact, the geometry deviates by a traceless factor from extremality which cannot be captured by  a conformal factor to AdS$_2$. To get an effective Schwarzian description, the deviations of both the metric and dilaton are equally important since they both grow similarly towards the boundary. The Schwarzian theory is one-loop exact, which reflects in the fact that we recover the same contribution from the large diffeomorphisms in a 4D one-loop computation. These two descriptions of near-extremal black holes are actually gauge-equivalent. In one description, the (tensor zero modes) fluctuations are realized from a bulk perspective in four dimensions whereas, in the 2D effective description, the fluctuations are localized on the near-horizon boundary.

We conclude this section with some important remarks that distinguish the above construction from that of the one presented in \cite{Nayak:2018qej, Moitra:2019bub, Iliesiu:2020qvm}.
It is well known that the Schwarzian theory appears as an effective description of Jackiw-Teitelboim (JT) gravity. In JT gravity, the large diffeomorphisms of AdS$_2$ acquire a Schwarzian action. 
Similarly, as we found above, the dynamics of near-extremal black holes can also be obtained from a Schwarzian  description that arises from the effective theory of  large diffeomorphisms on AdS$_2$. But there are interesting differences between the 4D Einstein-Maxwell theory around (near)extremality and JT gravity. In JT gravity, the background geometry is locally AdS$_2$, which is obtained by integrating out the dilaton field. On this geometry, the non-trivially varying dilaton captures the slight breaking of conformal invariance, giving rise to the Schwarzian theory. But in the case of a near-extremal black hole, the geometry is not locally AdS$_2$. The fluctuations of the geometry from AdS$_2$ appear in the same order as that of the fluctuations of the dilaton. These fluctuations of the geometry cannot be gauged away as is evident from the non-constancy of the Ricci scalar, even after taking care of the Weyl factor. Therefore, although the 1D Schwarzian description appears in both the gravity theories, the equivalence of Einstein-Maxwell theory around a near-extremal black hole and JT gravity is questionable. Nevertheless, the effective description of the large diffeomorphisms via a Schwarzian theory is manifest in both scenarios.

\section{Discussions} \label{summary-sec}
In this paper, we have studied the one-loop correction to the Euclidean partition function on a spherically symmetric electrically charged near-extremal background with charge $r_0$ and arbitrary small temperature $T$ in 4D Einstein-Maxwell theory. The quantum corrections are particularly important in the small temperature regime $r_0 T \ll 1$, where the semiclassical description is insufficient. In addition to the logarithm of area correction, the one-loop result contains a large contribution of the form $\log{T}$ which has been obtained from a Schwarzian effective action in \cite{Iliesiu:2020qvm, Iliesiu:2022onk}. We extract these $\log{T}$ corrections for a near-extremal black hole via direct computation of Euclidean path integral in 4D without referring to the effective lower-dimensional description. Along the line of standard procedure, we expand all the fields around their background solution and expand the action to quadratic order. Then the one-loop contribution can be obtained from the one-loop determinant of the kinetic operator i.e. from its eigenvalues.

In presence of a small temperature deviation, the infinite AdS$_2$ throat in the near-horizon geometry of an extremal black hole gets cut off at a finite yet very large distance. Hence, the quantum corrections in the near-horizon geometry are much larger than those coming from the asymptotic region of the near-extremal black hole, where it can be approximated by the full extremal geometry. We compute the one-loop determinant in this near-horizon region. We treat the near-horizon geometry of the near-extremal black hole as a linear order deviation from extremal AdS$_2\times$S$^2$ geometry, where the deviations are parametrized by the temperature. Because of this structure of the background, the near-extremal kinetic operator can be expressed as a small temperature correction to the extremal kinetic operator. Thereafter to evaluate the eigenvalues, we invoke the first-order perturbation theory. From this analysis, we understand that the origin of the $\log{T}$ contribution is due to the temperature-dependent mass acquired by the zero modes of the extremal operator in a near-extremal background. Contributions from other modes are polynomially suppressed in temperature and very small compared to the $\log{r_0}$ and $\log{T}$ contributions. We finally compute the total $\log{T}$ corrections coming from the tensor and $l = 1$ vector zero modes. In particular, the tensor mode contribution agrees with the Schwarzian result. 

Another important point to note is that the average thermodynamic energy and entropy can be computed as,
\begin{align}
    & \langle E \rangle = -\frac{\log{Z}}{\partial\beta} \sim E_{\text{cl}} + 3T, \\
    & \langle S \rangle = (1-\beta\partial_{\beta})\log{Z} \sim S_{\text{cl}} + 3\log{T}.
\end{align}

Here, $\beta$ is the inverse temperature parameter. We see that at very small temperature, the entropy approaches negative infinity and is unphysical\footnote{Similar issues have been raised in \cite{Almheiri:2014cka}.}. However, a non extremal black hole with any low temperature is certainly a physical object. To understand the issue better we find the density of states of the system \footnote{We thank Ashoke Sen for explaining this point to us.}. Since we are considering a spherically symmetric near-extremal black hole, we compute the density of states and entropy in a mixed ensemble (with fixed charge and energy), following \cite{Sen:2012dw},
\begin{align}
    \rho(E) = \int d\beta \text{e}^{\beta E}Z(\beta), \quad S(E) = \log{\rho(E)}.
\end{align}
Considering the logarithmic correction \eqref{full-logT} along with the semiclassical contribution above extremality, we have $Z(\beta) \sim \text{e}^{\frac{1}{\beta}}\beta^{-3}$. Therefore the density of states is given as,
\begin{align}
    \rho(E)\sim E J_2 (2\sqrt{E}) \xrightarrow{E\rightarrow 0} \frac{1}{2}E^2, 
\end{align}
here $J_\alpha (x)$ is the Bessel's function of first kind. Therefore, as the energy $E$ above extremality goes to zero, the density of states vanishes. At such low densities, the entropy is ill-defined and hence is not an appropriate physical quantity to look at. The system is perfectly well defined. We should note that this result of density of states will receive contributions from the $\mathcal{O}(T)$ corrections of the logarithm of the partition function. To understand the energy dependence of low-temperature density of states it is important to consider the temperature dependence appropriately. An advantage of our strategy of section \ref{pert-theory-sec} is that it paves a way to compute these $\mathcal{O}(T)$ corrections to near-extremal thermodynamics. On the contrary, it is very difficult to understand these corrections from a lower dimensional effective theory perspective, where we restrict only to the massless sector. The  $\mathcal{O}(T)$ computation would require keeping track of all the massive Kaluza-Klein modes. We would address the $\mathcal{O}(T)$ corrections in a future work.

Let us conclude the paper with some directions that can be explored further. Recently, localization in supersymmetric theories has been discussed in \cite{Iliesiu:2022kny, Sen:2023dps} for understanding the leading quantum corrections to the thermodynamics. It would be interesting to study the leading order quantum corrections in temperature for near-extremal partition function in such supersymmetric theories and to try to understand how much of these can be captured by (super)Schwarzian theories \cite{Heydeman:2020hhw, Iliesiu:2022kny}.  We would also like to address the question in a microscopic description of the black holes and try to see if similar corrections can be extracted from the microscopic side. In our earlier work \cite{Banerjee:2021vjy}, we studied the validity of the two-dimensional effective description of near-extremal black holes in a gravity theory perturbatively corrected by higher derivative interactions. In light of the present work, we understand that the effective description via a JT-like theory is questionable. Instead, we should be able to find the correct Schwarzian as described in section \ref{1D-sec}. We keep this check for our future study. 

\acknowledgments

We thank Ashoke Sen for numerous important discussions and suggestions on the work. We are thankful to Shamik Banerjee for discussions and collaborations at the initial stage of this work.  We are also thankful to   Suvankar Dutta, G. J. Turiaci and  V. Suneeta for helpful discussions and comments. NB would like to thank ICTS for its warm hospitality at an important stage of this work. MS would  like to thank Arindam Bhattacharjee, Debangshu Mukherjee and Gurmeet for useful discussions and comments. Finally, we would like to thank the people of India for their generous support towards research in basic sciences.

\appendix
\section{Basis for different fields and conventions} \label{basis-app}

For the sake of consistency, we will review the choice of basis on AdS$_2\times$S$^2$ for various fields. These are discussed in profound detail in \cite{Banerjee:2011jp, Sen:2012kpz, Sen:2012cj}. We will expand the fields in terms of the eigenfunctions of the Laplacian on AdS$_2$ and S$^2$. We will denote the four-dimensional coordinates as $x^A$, the coordinates on AdS$_2$ and S$^2$ as $x^\mu$ and $x^i$ respectively. Since both $AdS_2$ and $S^2$ are two-dimensional maximally symmetric spaces with characteristic radii $r_0$, we can write,
\begin{align}
    & R_{\mu\nu\rho\sigma} = \frac{R}{2}(g_{\mu\rho}g_{\nu\sigma} - g_{\mu\sigma}g_{\nu\rho}), \quad R_{\mu\nu} = \frac{R}{2}g_{\mu\nu}, \quad \text{with} \quad R = -\frac{2}{r_0^2} \\
    & R_{ijkl} = \frac{R}{2}(g_{ik}g_{jl} - g_{il}g_{jk}), \quad R_{ij} = \frac{R}{2}g_{ij}, \quad \text{with} \quad R = \frac{2}{r_0^2}
\end{align}
The gauge field strengths, being antisymmetric tensors in 2D, must be proportional to the Levi-Civita tensors. For our electrically charged extremal solution, we have
\begin{align}
    & \varepsilon_{\eta\theta} = r_0^2\sinh{\eta},\quad \varepsilon_{\psi\varphi} = r_0^2\sin{\psi} \\
    & F_{\mu\nu} = i\frac{Q}{r_0^2}\varepsilon_{\mu\nu},\quad F_{ij} = 0
\end{align}

\subsection*{Orthonormal basis in $AdS_2$}

\begin{itemize}
    \item Eigenfunctions of the Laplacian operator:
    \begin{align}
        & \nabla^2 W_p = -\hat{\kappa}_p W_p \label{scalar_eigen_AdS2}\\
        & \int_{AdS_2} W_p W_q = \delta_{pq} \label{scalar_ortho_AdS2}
    \end{align}
    \item Explicit expression for the eigenfunctions with the label ``$p$" representing $(\lambda,n)$ with $0<\lambda<\infty$ and $n\in \mathbb{Z}$,
    \begin{align}
        W_p \equiv f_{\lambda,n}(\eta,\theta) = & \frac{1}{\sqrt{2\pi r_0^2}}\frac{1}{2^{\abs{n}}\abs{n}!}\abs{\frac{\Gamma(i\lambda+\frac{1}{2}+\abs{n})}{\Gamma(i\lambda)}}\text{e}^{in\theta}\sinh^{\abs{n}}{\eta} \nonumber \\
        &F\left(i\lambda+\frac{1}{2}+\abs{n},-i\lambda+\frac{1}{2}+\abs{n}; \abs{n}+1; -\sinh^2{\frac{\eta}{2}}\right)
    \end{align}
    $F$ is the hypergeometric function. This has eigenvalue,
    \begin{align}
        \hat{\kappa}_p \equiv \frac{1}{r_0^2}\left(\lambda^2 + \frac{1}{4}\right) 
    \end{align}
    
    \item Normalized basis for vectors $\{\hat{\xi}^I_{p,\mu}\, , I=1,2\}$, which can be constructed out of the normalizable scalar eigenfunctions $W_p$. The ``$I$" label corresponds to the number of linearly independent vectors, the ``$p$" label characterizes the mode and the ``$\mu$" index is the vector index. Both the vectors have the same $\nabla^2$ eigenvalue.
    \begin{align}
        & \hat{\xi}^1_{p,\mu} = \frac{1}{\sqrt{\hat{\kappa}_p}}\nabla_\mu W_p, \quad \hat{\xi}^2_{p,\mu} = \frac{1}{\sqrt{\hat{\kappa}_p}}\varepsilon_{\mu\nu}\nabla^\nu W_p \label{vector_basis_AdS2}  \\
        & \nabla^2\hat{\xi}^I_{p,\mu} = -\left(\hat{\kappa}_p + \frac{1}{r_0^2}\right)\hat{\xi}^I_{p,\mu} \label{vector_eigen_AdS2} 
    \end{align}
    In addition to this, there are other normalizable vectors ${  v^I_{n,\mu},\, I=1,2}$ which are constructed out of derivatives acting on non-normalizable scalars on AdS$_2$, labeled by some discrete parameter `$n$'. These modes, corresponding to large gauge transformations have the following form,
    \begin{align}
        d\Phi_n, \quad \Phi_n \equiv \frac{1}{\sqrt{2\pi \abs{n}}}\left(\frac{\sinh{\eta}}{1+\cosh{\eta}}\right)^{\abs{n}}\text{e}^{in\theta}, \quad n = \pm 1,\pm 2\cdots \label{non-norm}
    \end{align}
    We construct a real basis for vectors by considering the real and imaginary parts of the vector in \eqref{non-norm}, which can be expressed as,
    \begin{align}
        & v^1_{n,\mu} \equiv v_{n,\mu}, \quad v^2_{n,\mu} \equiv \varepsilon_{\mu\nu}v^\nu_n \label{non-norm_basis_vector} \\
        & \nabla^2  v^I_{n,\mu} = -r_0^{-2} v^I_{n,\mu} \label{non-norm_eigen_vector} \\
        & \int g^{\mu\nu}\hat{\xi}^I_{p,\mu} \hat{\xi}^J_{q,\nu} = \delta^{IJ}\delta_{pq}, \quad \int g^{\mu\nu} v^I_{p,\mu} v^J_{q,\nu} = \delta^{IJ}\delta_{pq}, \quad \int g^{\mu\nu}\hat{\xi}^I_{p,\mu} v^J_{q,\nu} = 0 \label{vector_ortho_AdS2} 
    \end{align}
    Therefore any vector on AdS$_2$ must be expanded in the basis $\{\hat{\xi}^I_{p,\mu}, v^I_{p,\mu}\}$ for $I=1,2$, where the label `$p$' represents all the appropriate labels collectively in different categories.

    \item Normalized basis for symmetric rank two tensors $\{\hat{\chi}^P_{p,\mu\nu}\, , P=1,2,3\}$, which can be again constructed out of the scalar eigenfunctions $W_p$. The ``$P$" label corresponds to the number of linearly independent elements, the ``$p$" label characterizes the mode and the ``$\mu,\nu$" label are the tensor indices.
    \begin{align}
         & \hat{\chi}^I_{p,\mu\nu} = \frac{1}{\sqrt{\kappa_p+2r_0^{-2}}}(\nabla_\mu\hat{\xi}^I_{p,\nu}+\nabla_\nu\hat{\xi}^I_{p,\mu} - g_{\mu\nu} \nabla\cdot\hat{\xi}^I_p ), \quad \hat{\chi}^3_{p,\mu\nu} = \frac{1}{\sqrt{2}}g_{\mu\nu}W_p \label{tensor_basis_AdS2} \\
         & \nabla^2 \hat{\chi}^I_{p,\mu\nu} = - (\hat{\kappa}_p+4r_0^{-2})\hat{\chi}^I_{p,\mu\nu}, \quad \nabla^2\hat{\chi}^3_{p,\mu\nu} = -\hat{\kappa}_p\ \hat{\chi}^3_{p,\mu\nu} \label{tensor_eigen_AdS2} 
    \end{align}
    There are additional normalized tensor modes $w_{n,\mu\nu}$ corresponding to non-normalizable diffeomorphisms (or large diffeomorphisms), where $\{n, \, n = \pm 2,\pm 3 \cdots\}$ is a discrete label. These are given as,
    \begin{align}
         \frac{r_0}{\sqrt{\pi}}\left(\frac{\abs{n}(n^2-1)}{2}\right)^{1/2}\frac{(\sinh{\eta})^{\abs{n}-2}}{(1+\cosh{\eta})^{\abs{n}}}\text{e}^{in\theta}\left(d\eta^2+2i\sinh{\eta}d\eta d\theta - \sinh^2{\eta}d\theta^2\right) \label{non-norm_tensor}
    \end{align}
    These modes (constructed from the real and imaginary parts of \eqref{non-norm_tensor}) need to be added as linearly independent elements in the basis, which we denote as $\{\Omega^P_{p,\mu\nu}, P=1,2,3\}$ which are given as,
    \begin{align}
        & \Omega^I_{p,\mu\nu} = \frac{r_0}{\sqrt{2}}(\nabla_\mu v^I_{p,\nu}+\nabla_\nu v^I_{p,\mu} - g_{\mu\nu} \nabla\cdot  v^I_p ), \quad \Omega^3_{p,\mu\nu} \equiv w_{p,\mu\nu} \label{non-norm_basis_tensor} \\
        & \nabla^2\Omega^I_{p,\mu\nu} = -\frac{4}{r_0^2}\Omega^I_{p,\mu\nu}, \quad \nabla^2 w_{p,\mu\nu} = -\frac{2}{r_0^2}w_{p,\mu\nu} \label{non-norm_eigen_tensor} \\
        & \int g^{\mu\rho}g^{\nu\sigma}\hat{\chi}^P_{p,\mu\nu}\hat{\chi}^Q_{q,\rho\sigma} = \delta^{PQ}\delta_{pq}, \quad \int g^{\mu\rho}g^{\nu\sigma}\Omega^P_{p,\mu\nu}\Omega^Q_{q,\rho\sigma} = \delta^{PQ}\delta_{pq}, \nonumber \\
        &\int g^{\mu\rho}g^{\nu\sigma}\hat{\chi}^P_{p,\mu\nu}\Omega^Q_{q,\rho\sigma} = 0  \label{tensor_ortho_AdS2}
    \end{align}
     Therefore any symmetric rank two tensor on AdS$_2$ can be expanded in the basis $\{\hat{\chi}^P_{p,\mu},\Omega^P_{p,\mu}\}$ for $P=1,2,3$, where the label `$p$' represents all the appropriate labels collectively in different categories. 
\end{itemize}

\subsection*{Orthonormal basis in S$^2$}

\begin{itemize}
    \item Eigenfunctions of the Laplacian operator:
    \begin{align}
        & \nabla^2 U_p = -\kappa_p U_p \label{scalar_eigen_S2} \\
        & \int_{S^2} U_p U_q = \delta_{pq} \label{scalar_ortho_S2}
    \end{align}
    \item The explicit expression of the eigenfunctions and eigenvalues with the label ``$p$" representing $(l,m)$ for $l\in\mathbb{Z}^+$ and $-2l<m<2l$,
    \begin{align}
        U_p\equiv \frac{1}{r_0}Y_{lm}(\psi,\varphi) = \left(\frac{2l+1}{4\pi r_0^2}\frac{(l+\abs{m})!}{(l-\abs{m})!}\right)^{1/2}P^{-\abs{m}}_l (\cos{\psi})\text{e}^{i m \varphi}, \quad \kappa_p = \frac{l(l+1)}{r_0^2}
    \end{align}
    Here $Y_{lm}$ are the spherical harmonics.
    
    \item Normalized basis for vectors $\{\xi^I_{p,i}\, , I=1,2\}$, which can be constructed out of the scalar eigenfunctions $U_p$. The ``$I$" label corresponds to the number of linearly independent vectors, the ``$p$" label characterizes the mode and the ``$i$" label is the vector index. Both the vectors have the same $\nabla^2$ eigenvalues.
    \begin{align}
        & \xi^1_{p,i} = \frac{1}{\sqrt{\kappa_p}}\nabla_i U_p, \quad \xi^2_{p,i} = \frac{1}{\sqrt{\kappa_p}}\varepsilon_{ij}\nabla^j U_p \label{vector_basis_S2} \\
        & \nabla^2\xi^I_{p,i} = -\left(\kappa_p - \frac{1}{r_0^2}\right)\xi^I_{p,i} \label{vector_eigen_S2} \\
        & \int_{S^2} g^{ij}\xi^I_{p,i}\xi^J_{q,j}= \delta^{IJ}\delta_{pq} \label{vector_ortho_S2}
    \end{align}
    \item Normalized basis for symmetric rank two tensors $\{\chi^P_{p,ij}\, , P=1,2,3\}$, which can be again constructed out of the scalar eigenfunctions $U_p$. The ``$P$" label corresponds to the number of linearly independent elements, the ``$p$" label characterizes the mode, and the ``$i,j$" labels are the tensor indices.
    \begin{align}
         & \chi^I_{p,ij} = \frac{1}{\sqrt{\kappa_p-2r_0^{-2}}}(\nabla_i\xi^I_{p,j}+\nabla_j\xi^I_{p,i} - g_{ij} \nabla\cdot\xi^I_p), \quad \chi^3_{p,ij} = \frac{1}{\sqrt{2}}g_{ij}U_p \label{tensor_basis_S2} \\
         & \nabla^2 \chi^I_{p,ij} = - (\kappa_p-4r_0^{-2})\chi^I_{p,ij}, \quad \nabla^2\chi^3_{p,ij} = -\kappa_p\ \chi^3_{p,ij} \label{tensor_eigen_S2} \\
         & \int_{S^2} g^{ik}g^{jl}\chi^P_{p,ij}\chi^Q_{q,kl} = \delta^{PQ}\delta_{pq} \label{tensor_ortho_S2}
    \end{align}

\end{itemize}

\section{Semiclassical thermodynamics of Reissner-N\"{o}rdstrom solution} \label{RN-thermo-app}
In this section, we will review the computation of thermodynamic quantities of a Reissner-N\"{o}rdstrom black hole. Unlike the analysis of section \ref{ent-NHR-sec}, here we will take the boundary to infinity and perform background subtraction to regulate the action so that we have the correct expression for energy as well. Here, the form of the full geometry is required. The result for Bekenstein-Hawking entropy remains the same.

The regulated action is given as,
\begin{align}
    \mathcal{S} = -\int d^4x \sqrt{g} (R - F^2) -2\int_{r_\infty} d^3x \sqrt{\gamma} (K + 2 n_A A_B F^{AB}) + \frac{4}{r_\infty}\int_{r_\infty} d^3x \sqrt{\gamma} 
\end{align}
Here we have added a counterterm at the boundary which essentially regulates the energy by subtracting the contribution coming from flat space. The periodicity of the flat space is so chosen that asymptotically the it approaches the black hole geometry \cite{York:1986it}. In the computation of thermodynamic quantities, we will consider an ensemble where the charge and temperature are fixed.

\subsection{Non-extremal black hole}
To compute the thermodynamic quantities, we first compute the on-shell action for the non-extremal RN geometry. For the solution \eqref{RN}, we have:
\begin{align*}
    & n_A = \frac{1}{\sqrt{f(r_{\infty})}}(0,1,0,0), \quad \gamma_{ab} = \text{diag}(f(r_\infty),r_\infty^2,r_\infty^2\sin^2{\psi}) \\  
    & K = \frac{2}{r_\infty} - \frac{Q^2+r_{+}^2}{2r_{+}r_\infty^2} + \mathcal{O}\left(\frac{1}{r_\infty^3}\right), \quad A_B = i Q \left(\frac{1}{r_{+}}-\frac{1}{r}\right) (1,0,0,0)
\end{align*}
We find the regulated on-shell action as given by,
\begin{align}
    I = \frac{4\pi\beta}{r_{+}}(3Q^2 + r_{+}^2)
\end{align}
The energy for $r_\infty\rightarrow\infty$ is given as,
\begin{align}
    E = \frac{\partial I}{\partial \beta} = \frac{8\pi(Q^2+r_{+}^2)}{r_{+}} =16\pi M
\end{align}
The entropy is given by,
\begin{align}
    S_{\text{ent}} = \beta E - I = 16\pi^2 r_{+}^2
\end{align}
This is in agreement with Wald's formula \cite{Wald:1993nt}. It is worth noting, that the expression of entropy does not depend on the location of the boundary i.e. for this computation, the boundary can be put into any finite location. Neither does it depend on the counterterm.

\subsection{Near-extremal black hole}
In this subsection, we will compute the on-shell action for the near-extremal background and then compute the semiclassical contribution to the partition function and entropy. This result can be obtained by taking the small temperature limit of the computation for non-extremal black hole. But we will compute it from the near-horizon geometry and carefully consider the contributions coming from FHR. This analysis gives the correct expression for energy also. But for the computation of entropy, the near-horizon data is sufficient as in section \ref{ent-NHR-sec}.

The full geometry split into NHR and FHR as described in section \ref{NHR-sec}. We will consider the Einstein-Maxwell theory on these two manifolds separately. We add appropriate boundary terms and counterterm on the boundary $\partial M$ located at fixed radial distance $r = r_{\infty}$ near asymptotic infinity. For metric and gauge field, we impose Dirichlet and Neumann boundary conditions respectively. Now we split the action into two parts given as, $\mathcal{S} = \mathcal{S}_1 + \mathcal{S}_2$, such that:
\begin{align}
    & \mathcal{S}_1 = - \int_{\eta=0}^{\eta_0} d^4x \sqrt{g} (R - F^2) \label{act-near} \\
    & \mathcal{S}_2 = - \int_{r = r_b}^{r_\infty} d^4x \sqrt{g} (R - F^2) - 2\int_{\partial M} d^3x \sqrt{\mathfrak{h}} (K + 2 n_A A_B F^{AB}) + \frac{4}{r_\infty}\int_{\partial M} d^3x \sqrt{\mathfrak{h}} \label{act-far}
\end{align}
Here, the first part \eqref{act-near} of the action is evaluated on the NHR. We will see that the action \eqref{act-far} in the far part of the manifold generates a boundary term on the near-horizon boundary.

\subsubsection*{On-shell action in FHR:}
In FHR, the full near-extremal geometry is of the form $\{ g = \bar{g} +\delta g, A = \Bar{A}  + \delta A\}$, where $\{\Bar{g}, \Bar{A}\}$ denotes the full extremal geometry. Since the departure from extremality is very small, the on-shell action in the far part can be evaluated by plugging in the full near-extremal solution into \eqref{act-far},
\begin{align}
    I_{2}[g, A] = \mathcal{S}_{2}[\Bar{g}, \Bar{A}] + \delta\mathcal{S}_{2}
\end{align}

Since the extremal geometry also satisfies the equations of motion in FHR with periodicity of the time direction being $\beta$, the bulk part of the first-order variation term $\delta\mathcal{S}_{2}$ vanishes. From the bulk action, we have total derivative contributions that generate boundary terms on both the boundaries located at $r = r_b$ and $r = r_{\infty}$. Since $\delta g$ die off near infinity and $\delta F = 0$, the boundary terms generated at $r = r_{\infty}$ cancel with the Gibbons-Hawking and Maxwell boundary terms, consistent with the variational principle. Hence, we are left with a boundary term on the near-horizon boundary $r = r_b$. Therefore we have,
\begin{align}
    & \mathcal{S}_{2}[\bar{g},\bar{A}] = -32\pi^2 r_0^2 + \frac{64\pi^3 r_0^3}{\beta} - \frac{8\pi r_0}{r_b}(r_0 - 3r_b)\beta \nonumber \\
    & \delta\mathcal{S}_2 = -2\int_{\partial N} \sqrt{\mathfrak{h}}\left[(K + 2 n_A A_B F^{AB})_{\text{near-ext}} - (K + 2 n_A A_B F^{AB})_{\text{ext}}\right]
\end{align}
The normal on $\partial N$ points from the horizon to infinity. The on-shell action in far region is given as,
\begin{align}
    & I_{2}[g, A] = I_{\text{FHR}} -2\int_{\partial N} \sqrt{\mathfrak{h}}(K + 2 n_A A_B F^{AB})_{\text{near-ext}} \\
    &  I_{\text{FHR}} = 16\pi\beta \left(-r_0 + \frac{r_0^2}{r_b} + r_b\right)
\end{align}

This analysis shows that the geometry in the FHR can be well-approximated by the extremal geometry and it effectively generates a boundary term on the near-horizon boundary. We include this term in the NHR part of the action which is well-suited for the variational problem in this region. Supplementing the action \eqref{act-near} with this boundary term, we get:
\begin{align}
     & \mathcal{S}_{\text{NHR}} = -\int_{\eta=0}^{\eta_0} d^4x \sqrt{g} (R - F^2) -2\int_{\partial N} d^3x \sqrt{\mathfrak{h}} (K + 2 n_A A_B F^{AB}) \label{act-NHR}
\end{align}
As discussed earlier, the boundary $\partial N$ is located in the near-horizon region so that we consider it to be a small deviation from the horizon i.e. ${r_b = r_0(1+\varepsilon)}$ for $\varepsilon \ll 1$. Suppressing higher order terms in $\varepsilon$, we have:
\begin{align}
    I_{\text{FHR}} = 16\pi\beta r_0 (1+\varepsilon^2)
\end{align}
This is a divergent constant. As we will see below, the entire thermodynamics can be understood from the well-defined action \eqref{act-NHR} in the near-horizon region.

\subsubsection*{On-shell action in NHR:}
Now we plug in the near-horizon near-extremal solution given by \eqref{ext-NH} and \eqref{NE-corr} into the action \eqref{act-NHR} in NHR,
\begin{align}
    I_{\text{NHR}} = -16\pi^2 r_0^2 -\frac{32\pi^3 r_0^3}{\beta} \left(1 + \cosh{2\eta_0}\right) 
\end{align}
The location of the near-horizon boundary $\partial N$ is so chosen that it is asymptotically far from the horizon i.e. $\eta_0$ is large. But it should still remain in the near-horizon region with respect to the FHR geometry. This condition also imposes an upper bound on the near-horizon radial coordinate $\eta$. From \eqref{NH-coord} we have:
\begin{align}
    & r_b = r_{+} + \frac{2\pi r_0^2}{\beta} (\cosh{\eta_0}-1) \approx r_0(1+\varepsilon) \nonumber \\
    & \frac{\pi r_0}{\beta} \text{e}^{\eta_0} \simeq \varepsilon \ll 1 \label{NH-cutoff}
\end{align}
Therefore, the location of $\partial N$ is chosen such that the cutoff $\eta_0$ lies in the range,
\begin{align}
    1 \ll \text{e}^{\eta_0} \ll \frac{\beta}{r_0} \label{NHbdy-lim}
\end{align}
As we will now show that the physical results do not depend on this location as long as the boundary lies in this range. Using \eqref{NH-cutoff}, the on-shell action in NHR is given as,
\begin{align}
I_{\text{NHR}} = -16\pi^2 r_0^2 -\frac{32\pi^3 r_0^3}{\beta} - 16\pi\beta r_0 \varepsilon^2 \label{onshell-NHR}
\end{align}
We have suppressed the higher order terms in $\frac{1}{\beta}$ and $\varepsilon$.

\subsubsection*{Full on-shell action and semiclassical entropy}
The full on-shell action is given as,
\begin{align}
    I =  I_{\text{NHR}} +  I_{\text{FHR}} = -16\pi^2 r_0^2 -\frac{32\pi^3 r_0^3}{\beta} + 16\pi\beta r_0 \label{onshell}
\end{align}
The semiclassical partition function is given by $\log{Z_0} = - I$. The thermodynamic energy is given by,
\begin{align}
    E = \frac{\partial I}{\partial \beta} = 16\pi r_0 + \frac{32\pi^3 r_0^3}{\beta^2} \label{NE-energy}
\end{align}
This is equal to the mass parameter of the near-extremal solution given in \eqref{NE-parameters}. The entropy is given by,
\begin{align}
    S_{\text{ent}} = \beta E - I =   16\pi^2 r_0^2 \left(1 + \frac{4\pi r_0}{\beta} \right) \label{ent-cl}
\end{align}
This result is in agreement with the Bekenstein-Hawking entropy of the near-extremal black hole to order $\frac{1}{\beta}$.

\section{Solving the equations of motion in NHR} \label{NH-geo-app}

In order to understand the near-horizon geometry of the near-extremal black hole, we solve the equations of motion \eqref{eom} perturbatively in the near-horizon region of the black hole and recover the correct geometry obtained in section \ref{NHR-sec} from the full solution. The near-horizon geometry is a small deviation from the extremal one of the form: $g_{AB} = \bar{g}_{AB} + \tilde{\epsilon} g^{(c)}_{AB}, F_{AB} = \bar{F}_{AB} + \tilde{\epsilon} F^{(c)}_{AB}$ i.e. the unperturbed solution is of the form AdS$_2\times$S$^2$,
\begin{align}
        & \bar{g}_{AB}dx^A dx^B = r_0^2 (d\eta^2 + \sinh^2{\eta}d\theta^2) + r_0^2 (d\psi^2 + \sin^2{\psi}d\varphi^2) , \quad \bar{F}_{\mu\nu} = \frac{i}{r_0}\varepsilon_{\mu\nu}
\end{align}
Here $\varepsilon_{\mu\nu}$ is the Levi-Civita tensor on $AdS_2$, with the non-zero component being ${\varepsilon_{\eta\theta} = r_0^2 \sinh{\eta}}$. The perturbation parameter $\tilde{\epsilon}$ is to be determined by matching the geometry with the full solution. Now we consider the near-extremal correction to the extremal background \eqref{ext-NH} to be of the following form,
\begin{align}
    g^{(c)}_{AB}dx^A dx^B = &\chi (x^\mu) r_0^2 (d\eta^2 + \sinh^2{\eta}d\theta^2) + \alpha (x^\mu) r_0^2(d\eta^2 - \sinh^2{\eta}d\theta^2) \nonumber \\
    & + \phi(x^\mu) r_0^2(d\psi^2 + \sin^2{\psi}d\varphi^2); \quad F^{(c)}_{\mu\nu} = \frac{i}{r_0}\Theta (x^\mu) \varepsilon_{\mu\nu} \label{NE-ansatz} 
\end{align}
Solving the equations of motion up to order $\tilde{\epsilon}$, we get the following solutions of the parameters appearing in the ansatz \eqref{NE-ansatz},
\begin{itemize}
    \item \subsubsection*{Branch-1: Fluctuating AdS$_2$ radius and gauge field strength}
    \begin{align}
        \chi(\eta) = c_2 + \cosh{\eta}(c_1 - c_2 \tanh^{-1}{(\sech{\eta})}); \quad \alpha(\eta) = 0; \quad \phi(\eta) = 0; \quad \Theta(\eta) = \chi(\eta) \label{NE-corr1}
    \end{align}
    The small $\eta$ expansion of the solution is given as,
    \begin{align}
        \chi(\eta)\xrightarrow{\eta\rightarrow 0} c_1 + c_2 + c_2\ln{\frac{\eta}{2}} + \frac{\eta^2}{12}(6c_1 - c_2 + 6c_2\ln{\frac{\eta}{2}} )
    \end{align}
    Imposing regularity at $\eta = 0$, we set $c_2 =0$.
    \item \subsubsection*{Branch-2: Traceless fluctuation on AdS$_2$}
     \begin{align}
        \chi(\eta) = 0; \quad \alpha(\eta) = a_2 + \coth^2{\eta}(-a_2 + a_1\sech{\eta}); \quad \phi(\eta) = 0; \quad \Theta(\eta) = 0; \label{NE-corr2}
    \end{align}
    We consider the small $\eta$ expansion of $\alpha(\eta)$,
    \begin{align}
        \alpha(\eta)\xrightarrow{\eta\rightarrow 0} \frac{a_1 - a_2}{\eta^2} + \frac{1}{6}(a_1 + 2a_2) + \frac{1}{120}(-7a_1-8a_2)\eta^2
    \end{align}
    We set $a_1 = a_2$ so that the solution does not blow up at $\eta = 0$, then we have:
    \begin{align}
         \alpha(\eta)\xrightarrow{\eta\rightarrow 0} \frac{a_1}{2}\left(1 - \frac{\eta^2}{4}\right)
    \end{align}
    \item \subsubsection*{Branch-3: Traceless fluctuation on AdS$_2$, fluctuating S$^2$ radius and gauge field strength}
     \begin{align}
        &\alpha(\eta) = \frac{1}{2}\coth{\eta}\csch{\eta}(1 + 2b_1 + \cosh{2\eta} - 2b_2\sech{\eta}); \nonumber\\
        &\chi(\eta) = 0; \quad \phi(\eta) = \cosh{\eta}; \quad \Theta(\eta) = -\cosh{\eta} \label{NE-corr3}
    \end{align}
    We study the behavior of these fluctuations near $\eta\rightarrow 0$.
    \begin{align}
         \alpha(\eta)\xrightarrow{\eta\rightarrow 0} \frac{1+b_1 - b_2}{\eta^2} + \frac{1}{6}(7+b_1+2b_2)+\frac{1}{120}(53-7k_1-8k_2)\eta^2
    \end{align}
    From the demand that it does not blow up at $\eta = 0$, we get $b_2 = 1 + b_1$ such that,
    \begin{align}
         \alpha(\eta)\xrightarrow{\eta\rightarrow 0} \frac{3+b_1}{2} + \frac{1}{8}(3-b_1)\eta^2
    \end{align}
    If we further demand that $\gamma_{\mu\nu}\rightarrow 0$ as $\eta\rightarrow 0$, we get $b_1 = -3$ such that,
    \begin{align}
        \alpha(\eta) = (2 + \cosh{\eta})\tanh^2{\left(\frac{\eta}{2}\right)}
    \end{align}
    On the horizon i.e. at $\eta = 0$, the time component of metric should go to zero. Under this demand, the first two branches of solutions are identically zero. 
\end{itemize}
Therefore, the near-extremal deviation \eqref{NE-ansatz} in the near-horizon region is given as,
\begin{align}
    g^{(c)}_{AB}dx^A dx^B =  & (2 + \cosh{\eta})\tanh^2{\left(\frac{\eta}{2}\right)}r_0^2(d\eta^2 - \sinh^2{\eta}d\theta^2)\nonumber \\
    & + \cosh{\eta}\ r_0^2(d\psi^2 + \sin^2{\psi}d\varphi^2); \quad F^{(c)}_{\mu\nu} = -\frac{i}{r_0} \cosh{\eta}\ \varepsilon_{\mu\nu}
\end{align}
This is the same geometry \eqref{NE-corr} that we obtained from the full near-extremal solution with the identification $\tilde{\epsilon} = \frac{2\delta}{r_0} = 4\pi r_0 T$. Therefore, we conclude that the near-horizon geometry discussed in section \ref{NHR-sec}, is the unique spherically symmetric solution of the equations of motion to order $T$ in the NHR.

\bibliography{NElog.bib}
\end{document}